\documentclass[useAMS,usenatbib]{mn2e} 

\usepackage{psfig}
\usepackage{graphicx}
\usepackage{aas_macros}
\usepackage{epsf}
\usepackage{graphics}
\usepackage{amssymb}   
\usepackage{setspace}  

\newlength{\plotwidth}
\newlength{\fullwidth}
\newcommand{\persqdeg}{\mbox{\,deg$^{-2}$}}
\newcommand{\permag}{\mbox{\,mag$^{-1}$}}
\newcommand{\magdeg}{\mbox{\,\permag\persqdeg}}
\newcommand{\sqdeg}{\mbox{\,deg$^{2}$}}

\newcommand{\invcubicMpc}{\mbox{$\,h^3\,{\rm Mpc}^{-3}$}}
\newcommand{\kms}{\mbox{\,km\,s$^{-1}$}}

\newcommand{\aimsixdf}{150\,000}

\newcommand{\kb}{\mbox{$K$}}
\newcommand{\hb}{\mbox{$H$}}
\newcommand{\jb}{\mbox{$J$}}
\newcommand{\bj}{\mbox{$b_{\rm\scriptscriptstyle J}$}}
\newcommand{\rf}{\mbox{$r_{\rm\scriptscriptstyle F}$}}

\newcommand{\dkb}{\mbox{$\Delta K$}}

\newcommand{\khjrb}{\mbox{$KHJr_{\rm\scriptscriptstyle F}b_{\rm\scriptscriptstyle J}$}}
\newcommand{\brjhk}{\mbox{$b_{\rm\scriptscriptstyle J}r_{\rm\scriptscriptstyle F}JHK$}}
\newcommand{\jhk}{\mbox{$JHK$}}
\newcommand{\khj}{\mbox{$KHJ$}}
\newcommand{\br}{\mbox{$b_{\rm\scriptscriptstyle J}r_{\rm\scriptscriptstyle F}$}}

\newcommand{\ktot}{\mbox{$K_{\rm tot}$}}
\newcommand{\kiso}{\mbox{$K_{\rm iso}$}}
\newcommand{\mk}{\mbox{$M_K$}}
\newcommand{\mh}{\mbox{$M_H$}}
\newcommand{\mj}{\mbox{$M_J$}}

\newcommand{\al}{\mbox{$\alpha$}}
\newcommand{\ms}{\mbox{$M_*$}}
\newcommand{\ps}{\mbox{$\phi_*$}}

\newcommand{\pc}{\mbox{$\phi_{\rm conv}$}}

\newcommand{\lvmax}{\mbox{$\log[\phi_{\rm vmax}]$}}
\newcommand{\lswml}{\mbox{$\log[\phi_{\rm swml}]$}}

\newcommand{\vmax}{\mbox{$1/V_{\rm max}$}}

\newcommand{\plotone}[1]
    {\centering \leavevmode \psfig{file=#1,width=\plotwidth,clip=}}

\newcommand{\plotfull}[2]
    {\centering \leavevmode \psfig{file=#1,width=#2\fullwidth,clip=}}

\title[Luminosity Functions from 6dFGS] 
{Near-Infrared and Optical Luminosity Functions \\ from the 6dF Galaxy Survey}

\author[Jones et al.]{
\parbox[t]{\textwidth}{
D.~Heath Jones$^{1,2}$, Bruce A.~Peterson$^{2}$, Matthew Colless$^{1}$, 
Will Saunders$^{1}$\\
}
\vspace*{6pt} \\
$^1$ Anglo-Australian Observatory, P.O.\ Box 296, Epping, NSW 1710,
Australia\\ ({\tt heath@aao.gov.au, colless@aao.gov.au, will@aao.gov.au})\\
$^2$ Research School of Astronomy \&; Astrophysics, The Australian
National University, \\ Weston Creek, ACT 2611, Australia
({\tt peterson@mso.anu.edu.au})\\
} \date{Accepted ---. Received ---; in original form ---.}

\begin{document}

\maketitle

\begin{abstract}
  
  Luminosity functions and their integrated luminosity densities are
  presented for the 6dF Galaxy Survey (6dFGS). This ongoing survey
  ultimately aims to measure around 150\,000 redshifts and 15\,000
  peculiar velocities over almost the entire southern sky at
  $|b|>10^\circ$. The main target samples are taken from the 2MASS
  Extended Source Catalog and the SuperCOSMOS Sky Survey catalogue, and
  comprise 138\,226 galaxies complete to (\kb,\hb,\jb,\rf,\bj) = (12.75,
  13.00, 13.75, 15.60, 16.75). These samples are comparable in size to
  the optically-selected Sloan Digital Sky Survey and 2dF Galaxy
  Redshift Survey samples, and improve on recent near-infrared-selected
  redshift surveys by more than an order of magnitude in both number and
  sky coverage. The partial samples used in this paper contain a little
  over half of the total sample in each band and are $\sim$90\,percent
  complete.
  
  Luminosity distributions are derived using the \vmax, STY and SWML
  estimators, and probe 1 to 2 absolute magnitudes fainter in the
  near-infrared than previous surveys. The effects of magnitude errors,
  redshift incompleteness and peculiar velocities have been taken into
  account and corrected throughout. Generally, the 6dFGS luminosity
  functions are in excellent agreement with those of similarly-sized
  surveys. Our data are of sufficient quality to demonstrate that a
  Schechter function is not an ideal fit to the true luminosity
  distribution, due to its inability to simultaneously match the faint
  end slope and rapid bright end decline. Integrated luminosity
  densities from the 6dFGS are consistent with an old stellar population
  and moderately declining star formation rate.

\end{abstract}




\section{Introduction}
\label{sec:introduction}

Arguably the most fundamental of all cosmological observables is the
mean space density of galaxies per unit luminosity, or luminosity
function (LF). In the local universe, the shape of the LF places strong
constraints on models of dark halo formation and galaxy evolution.
Current semianalytic models that invoke feedback in the form of
supernovae and stellar winds are unable to satisfactorily reproduce the
sharp bright-end decline, while hydrodynamic simulations typically
produce LFs that are too bright \citep[e.g.][and references
therein]{benson03,kay02}. 
Alternative approaches \citep[e.g.][]{cooray05,berlind02,yang03} use
the LF as an empirical constraint
when inferring  how galaxies populate dark matter halos, thereby providing
testable predictions about galaxy clustering.

In the near-infrared (NIR), LFs are effective tracers of the stellar
mass function for many reasons. Unlike the visible or far-infrared,
light from the NIR is dominated by the older and cooler stars that make
up the bulk of the stellar mass. Compared to a survey at visible
wavelengths, the balance of galaxy types in a NIR survey shifts from
late to early types, which contain most of the stellar mass of the
universe. Furthermore, the effects of extinction are minimal at longer
wavelengths, meaning that NIR luminosities are affected little by dust,
either in our own Galaxy or in the target galaxy. Mass-to-light ratios
are also much better constrained in the NIR passbands \citep{bell01}.

Our current understanding of the galaxy LF owes much to the 2dF Galaxy
Redshift Survey
\citep[2dFGRS;][]{folkes99,cross01,norberg02,madgwick02,depropris03,eke04,croton05}
and the Sloan Digital Sky Survey
\citep[SDSS;][]{blanton01,goto02,blanton03,baldry05,berlind05,blanton05}.
The wide sky coverage and extremely large optically-selected samples
($\sim 10^5$) utilised in these studies have produced a tenfold increase
in the precision with which the shape of the LF is known. With the
advent of the Two-Micron All-Sky Survey \citep[2MASS;][]{jarrett00},
other studies \citep{cole01,kochanek01,bell03, eke05} have combined the 2MASS
NIR galaxy samples with the 2dFGRS, SDSS and ZCAT \citep{huchra92}
redshift catalogues to yield wide sky coverage and moderately large
NIR-selected samples ($\sim 10^4$).

In this paper we present \khjrb\ LFs from an interim subset of the 6dF
Galaxy Survey \citep[6dFGS;][]{jones04,jones05}, a NIR-selected redshift
survey of \aimsixdf\ galaxies over almost the entire southern sky
combined with a 15\,000-galaxy peculiar velocity survey. The full target
list contains 174\,442 sources, although most of these (138\,226) come
from five main samples that are complete to $(\kb,\hb,\jb,\rf,\bj) =
(12.75,13.00,13.75,15.60,16.75)$. The remainder are provided by a dozen
supporting samples selected in various ways \citep{jones04}. When
complete, the 6dFGS will cover a sky area eight times that of 2dFGRS and
twice that of the revised SDSS. The median 6dFGS redshift ($\bar{z} =
0.054$) is roughly half of that of the other two surveys, and reflects
its shallower limiting magnitudes. At completion, the 6dFGS will be
two-thirds the sample size of 2dFGRS and about one-fifth of SDSS.
\citet{jones04,jones05} give full descriptions of the survey
and show redshift maps of the galaxy distributions obtained thus far.

As well as providing initial results on the 6dFGS LFs, this paper gives
a comprehensive explanation of the LF derivations used in this and
related future papers. Section~2 summarises the target selection, galaxy
photometry and corrections for incompleteness; surface brightness
selection issues are also discussed. Section~3 examines normalisation
issues via number count data, and explains how we have accounted for
galaxy magnitude errors, peculiar velocities, extinctions and
$k$-corrections. In Section~4 we present the 6dFGS LFs and compare the
results to other recent work. Section~5 is devoted to a comparison of
the 6dFGS luminosity density across \brjhk\ with other relevant surveys.

We have adopted the current standard cosmology ($\Omega_{\rm M} = 0.3,
\Omega_{\Lambda} = 0.7$) throughout the paper.


\begin{figure*}
\plotfull{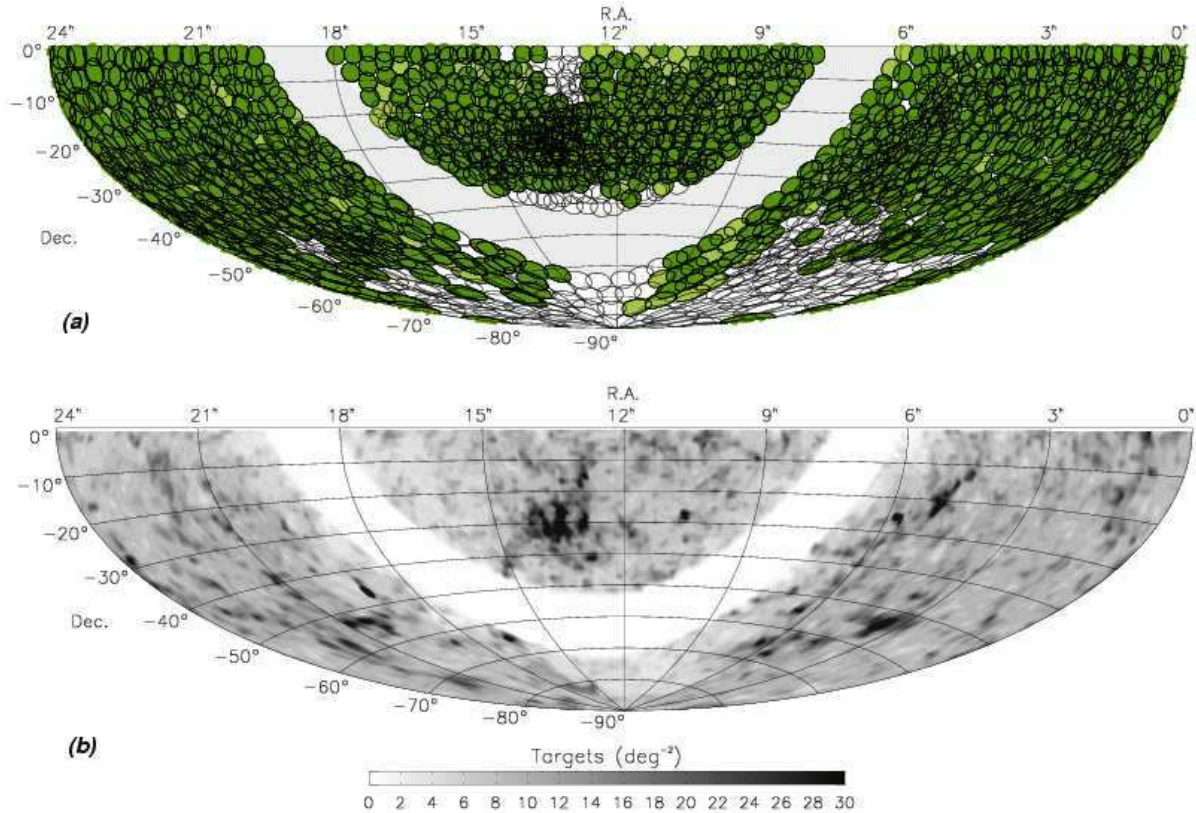}{0.9}
\caption{
  (a)~Equal area Aitoff projection of the total 6dFGS field coverage
  (circles) superposed with the subset of fields contributing to this
  paper (filled circles). (b)~Galaxy density (per square degree) of
  6dFGS targets on the sky.}
\label{fig:skycoverage}
\end{figure*}

\section{Target Selection and Photometry}
\label{sec:data}

\subsection{Areal and Magnitude Selection}

The 6dF Galaxy Survey commenced observations in 2001 May although formal
observing did not start until 2002 January. Incremental public data
releases were made in 2002 December, 2004 March, and 2005 May, the
latter making available on-line 83\,014 sources with their spectra,
redshifts and photometric measurements. \citet{jones04} describe the
First Data Release (2004 March) and the characteristics of the 6dFGS in
general, while \citet{jones05} describe the updates in place for the
Second Data Release (2005 May).

In total, there are 1564 fields and 174\,442 galaxies in the 6dFGS,
yielding a mean source density of 112 galaxies per field. With an
instrument field size of 25.5\,\sqdeg\ and 10.2 sources \persqdeg, the
field tiling has effectively had to cover the southern sky twice over.
Figure~\ref{fig:skycoverage} shows the survey coverage to date from the
1\,390 fields contributing to the LFs herein. Although substantial areas
at mid- and equatorial declinations have been covered, some areas await
a second pass, and as a result, redshift completeness will ultimately be
higher. Coverage around the Pole is still largely incomplete, but will
be addressed during the closing stages of the survey.

Targets for the 6dFGS were selected from the 2MASS Extended Source
Catalog \citep[2MASS XSC;][]{jarrett00} for southern declinations
$\delta < 0^\circ$ and at galactic latitudes $|\,b\,| > 10^\circ$ for
\jhk\ and $|\,b\,| > 20^\circ$ for \br. Table~\ref{tab:targets}
summarises the selection criteria, sample sizes and sky coverage. The
motivations for the sample selection criteria are outlined in
\citet{jones04}. 

Total magnitudes are the preferred way of quantifying galaxy flux
because the related observable (total luminosity) is more physically
meaningful than isophotal or aperture magnitudes that depend on
selection criteria or distance. At the time the target lists were
prepared, however, the total \kb\ magnitudes (\ktot) from the 2MASS XSC
were not sufficiently reliable at the lowest galactic latitudes in the
survey due to insufficient depth and spatial resolution. The isophotal
2MASS \kb\ magnitudes (\kiso) were more robust, however, and so formed
the basis for a {\sl derived} \ktot\ that we used for target selection
in \kb:
\begin{equation}
  K_{\rm tot} = K_{\rm iso} - 1.5 \exp{[1.25 (\overline{\mu_{K20}}-20)]} .
\label{magcorrn}
\end{equation}
Here, $\overline{\mu_{K20}}$ is the mean surface brightness within the
$\mu_K=20$ elliptical isophote. The scatter in $(\kiso\ - \ktot)$ is
typically no more than 0.1 to 0.2~magnitudes over the range of
surface-brightnesses encountered \citep[Fig.~1 of][]{jones04}.

This correction was eventually superseded by the revised total
\kb\ magnitudes from 2MASS XSC which became available after 6dFGS
observations had commenced. These show less scatter than the
\ktot\ values used for the 6dFGS sample selection (T.~Jarrett, priv.\ 
comm.) and are a superior measure of total flux.

Both the old and new magnitudes contribute to the 6dFGS luminosity 
functions in different ways. The original \ktot\ magnitudes are preserved
for selection purposes and were used for deriving the selection function 
and completeness corrections (Sect.\ \ref{sec:selection}). However, the superior 
\kb\ magnitudes are used to compute galaxy luminosities after
all selection cuts and completeness corrections have been made.

The first studies to use 2MASS photometry for LF work were divided over
the use of isophotal \citep{kochanek01} or total magnitudes
\citep{cole01}. While \citeauthor{kochanek01} claimed their $(\kiso -
\ktot)$ correction was only 0.20\,mag, \citet{andreon02} argued that it
is closer to 0.3\,mag. Claims that differences between the optical SDSS
and 2MASS NIR luminosity densities were due to a 2MASS bias against low
surface brightness systems \citep{wright01} were largely countered when
the SDSS optical luminosities were re-computed to include galaxy
evolution \citep{blanton03}, although a small offset remains.

\citet{cole01} have undertaken a comparison of their adopted 2MASS Kron
and extrapolated magnitudes with the deeper infrared photometry of
\citet{loveday00}. They found scatter of around $\sigma \approx
0.13$--0.14 magnitudes in both samples, although found that this reduced
to $\sigma \approx 0.1$ if \kb\ Kron magnitudes were derived from higher
quality \jb\ data and $(J-K)$ colours. Likewise, \citet{bell03} find a
scatter if $\sim 0.2$\,mag in comparisons of 2MASS Kron magnitudes with
those of \citet{loveday00}. Since these studies, the 2MASS photometry
has been revised. With the current total magnitudes, we find 
$\langle\kiso-\ktot\rangle=0.08$ for $\kiso \lesssim 10$, rising to
$\langle\kiso-\ktot\rangle=0.12$. At $\kiso \gtrsim 11$, the corrections
can be as large as 0.5\,mag, reflecting the lower surface brightness
late-type galaxies that populate this regime. \citet{eke05} find that the
improved 2MASS total magnitudes match the \citeauthor{loveday00} data
with a scatter of 0.125\,mag.

The original \br\ samples for 6dFGS were selected from the SuperCOSMOS
catalogue in 2001. This photometry suffered from scatter in the
field-to-field zeropoints of a few tenths of a magnitude.  After the
all-sky digital photometry of 2MASS subsequently became available, the
original \bj, \rf\ (and $i_{\rm N}$) SuperCOSMOS magnitudes were
re-calibrated in 2003 to a common zeropoint by J.~Peacock, N.~Hambly and
M.~Read, for the benefit of the
2dF Galaxy Redshift Survey.
Individual field zeropoints were characterised by the mean $(\bj\ - J)$
colour. The revised SuperCOSMOS magnitudes show a significant
improvement in the photometric calibration, with zeropoint scatter
of 0.04\,mag rms in each band \citep{cole05}.
\citet{cross04} find a scatter of $\sigma = 0.1$~mag in a comparison of
the revised SuperCOSMOS magnitudes with the CCD photometry of the
Millennium Galaxy Catalogue and Sloan Early and First Data Releases
\citep{stoughton02,abazajian03}. 

As with the revised \kb\ photometry, the original \br\ magnitudes were
retained for selection and completeness calculations while the re-calibrated
ones were used for luminosities. This is similar to the approach used 
by the 2dFGRS when re-calibrated \bj\ photometry became available mid-way through 
that survey \citep{norberg02,cole05}. However, in that case, the new and
old magnitudes were related by a simple transformation which imparted no 
additional scatter. In the case of 6dFGS, the relationship between the old and 
new magnitudes is not as straightforward. However, the scatter that results 
is small and inconsequential, largely because the dataset itself 
remains the same.

\begin{table*}
\begin{center}
\caption{6dFGS target samples\label{tab:targets}
} 
\vspace{6pt}
\begin{tabular}{l@{\hspace{12mm}}ccc@{\hspace{12mm}}cc}
\hline \hline
Sample Selection        &  \multicolumn{3}{c}{{\bf Full Sample}} & \multicolumn{2}{c}{{\bf This Paper}} \\
                        &  Total        & Additional    &  Area         &  Number  &  Effective Area \\
                        &  Number       & to \kb\ sample&  (deg$^2$)    &  Used    &  (deg$^2$) \\
\hline
2MASS $K<12.75$         & 113\,988      &  --           & 17\,045       & 60\,689  &  9\,075    \\
2MASS $H  <13.00$       &  90\,407      & 3\,282        & 17\,045       & 53\,142  & 10\,019    \\
2MASS $J  <13.75$       &  93\,925      & 2\,008        & 17\,045       & 55\,236  & 10\,024    \\
SuperCOSMOS $\rf<15.60$ &  66\,904      & 9\,199        & 13\,572       & 37\,156  &  7\,537    \\
SuperCOSMOS $\bj<16.75$ &  64\,138      & 9\,749        & 13\,572       & 37\,128  &  7\,857    \\
\hline
\end{tabular}
\end{center}
\end{table*}

\subsection{Surface Brightness Issues}

One common misgiving surrounding apparent magnitude limited surveys is
the extent to which they bias the LF against low surface brightness
systems \citep{disney76,impey97, sprayberry97,dalcanton98}.  Analyses
utilising bivariate brightness distributions find $\lesssim
1$~mag~arcsec$^{-2}$ change in the peak surface brightness for the
\bj-selected 2dF Galaxy Redshift Survey across observed luminosities
\citep{cross01,driver04}, while \citet{cross02} claim that the faint-end
slope is robust to surface brightness effects.

In the case of the 2MASS XSC, \citet{cole01} have modelled the selection
characteristics following the methods of \citet{cross01} and
\citet{phillipps90}, and find no selection bias against low surface
brightness galaxies by 2MASS. Specifically, they demonstrate that their
galaxies lie well within the theoretical maximum volumes corresponding
to their apparent magnitude and surface brightness limits ($\jb = 14.45$
and 20.5~mag~arcsec$^{-2}$ respectively). The claims by \citet{bell03}
that surface brightness selection impacts the \kb-band LF are based on
brighter surface brightness limits ($\mu_K = 17$ to 20) than the $\mu_K
= 20$ limit of the 6dFGS.

Taken together, these results imply that at the magnitudes used
by the 6dFGS, 2MASS is virtually unaffected by bias against low surface 
brightness galaxies . The 6dFGS magnitude limits are $\sim 1.5$~mag brighter
than the incompleteness regime for the 2MASS XSC, ensuring an effective
compromise between sensitivity and robust selection for the redshift
survey sample.

\subsection{Selection Function}
\label{sec:selection}

The probability of obtaining a redshift for a galaxy is expected to
depend on its apparent magnitude $K$: the fainter it is, the less the
chance that its redshift will be secured. However, in a survey such as
6dFGS, the basic observable units are not the galaxies, but the 
{\sl fields} to which they belong. Therefore, the probability that a
galaxy redshift will be obtained depends not only on its apparent
magnitude, but also on the overall redshift success rate of its field.

The survey selection function gives the survey redshift completeness as
function of both magnitude and sky position. In short, it is obtained by
taking each 6dFGS field and fitting a functional form to the decline in
completeness with magnitude \citep{colless01,norberg02}.  This form is
then normalised to reproduce the overall redshift completeness at that
point in the sky.

\begin{figure}
\plotone{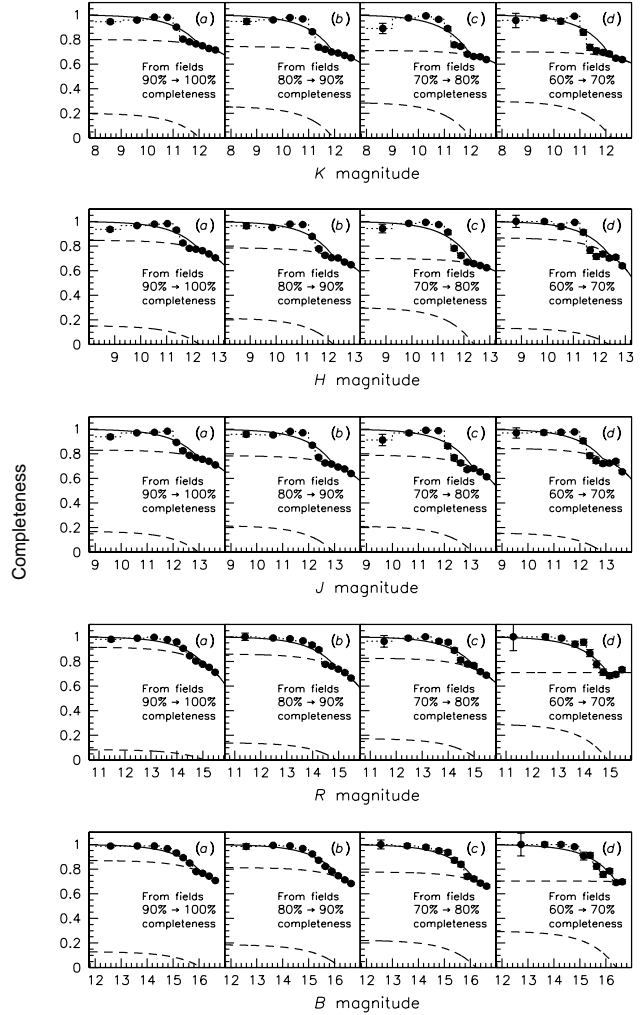}
\caption{
  Generalised field completeness and its dependence on apparent
  magnitude (solid points with error bars) for fields in the ranges
  ($a$) $0.9 < R_f \leq 1.0$, (set $F=1$), ($b$) $0.8 < R_f \leq 0.9$,
  ($F=2$), ($c$) $0.7 < R_f \leq 0.8$, ($F=3$), and ($d$) $0.6 < R_f
  \leq 0.7$, ($F=4$). The magnitude bin widths (dotted histogram) and
  individual fit components (dashed curves) are also shown.}
\label{fig:magcompl}
\end{figure}

Before we can compute the survey selection function we must first
determine the general form for completeness as a function of magnitude
for each band \khjrb, modulo field completeness. Figure
\ref{fig:magcompl} shows how this varies if we divide galaxies into one
of four groups:
\begin{enumerate}
\item galaxies from high quality fields with completeness in the range 
      $0.9 < R_f \leq 1.0$ (denoted set $F=1$);
\item galaxies from lower quality fields with $0.8 < R_f \leq 0.9$ ($F=2$); 
\item galaxies from mediocre fields with $0.7 < R_f \leq 0.8$ ($F=3$); 
\item galaxies from poor fields with $0.6 < R_f \leq 0.7$ ($F=4$). 
\end{enumerate}

The completeness, $R(f)$, of field $f$ was determined as
\begin{eqnarray}
  R(f) &  = & \frac{N_z(f)}{N_p(f)}  \nonumber\\
       &  = & \frac{ N_{{\rm lit}}(f) + N_{{\rm 6dF}}(f) }
       { N_{{\rm lit}}(f) + N_{{\rm 6dF}}(f) + N_{\rm f}(f) + N_{\rm r}(f) } .
\label{fieldcompl}
\end{eqnarray}
Here, $N_z(f)$ is the number of extragalactic redshifts obtained in the
field, $f$, and $N_p(f)$ is the number of confirmed-extragalactic and
yet-to-be-observed sources from the parent catalogue. $N_{\rm f}(f)$ and
$N_{\rm r}(f)$ are the number of redshift failures and sources remaining
to be observed, respectively. $N_{{\rm lit}}(f)$ and $N_{{\rm 6dF}}(f)$
count the literature and 6dF-measured redshifts within each field
boundary, respectively. Around one-eighth of our redshifts come from the
literature, mainly Huchra's ZCAT Catalog
\citep{huchra92} or the 2dFGRS \citep{colless01} in the ratio $4:1$
respectively. While observing for 2dFGRS is complete, ZCAT continues to
be supplemented by the Two Micron All-Sky Redshift Survey
\citep[2MRS;][]{huchra06}, which to date has obtained around 20\,k
redshifts to $\kb = 11.25$ across the full sky. Sources for which
spectra were obtained, and which were found to be stars, planetary
nebulae, H{\sc ii} regions, or other Galactic objects, were excluded.

At  $\bj \lesssim 16.0$ ($\kb \lesssim 12$), ZCAT redshifts represent more than 
one-third of the sample. As a consequence, completeness below the 2MRS limit 
is comparatively high while the 6dFGS continues to fill out completeness
at fainter magnitudes ($\kb \gtrsim 11.25$). This dual redshift contribution 
is evident in our measurements of completeness as a function of magnitude, 
which exhibit a bump in the declining portion of the curve  
(Fig.~\ref{fig:magcompl}).

The generalised completeness curves were fit by a double-exponential,
\begin{eqnarray}
  C_F(m) & = & \beta_F \,{\rm max}[0, 1-\exp{(m-\mu_F)}] \nonumber\\ 
         &   &  + (1-\beta_F)\,{\rm max}[0, 1-\exp{(m-\xi_F})] .
\label{doubleexp}
\end{eqnarray}
The motivation for this model was the single exponential form adopted
by 2dFGRS \citep{colless01},  modified to a double-exponential
to fit the shoulder.
Table~\ref{tab:parameters} lists the fit parameters $\beta_F$, $\mu_F$
and $\xi_F$ according to passband and field group ($F = 1,2,3$ or 4).
The shoulder occurs at $\xi_F$ values typically 0.5 to 0.75\,mag 
brighter than the survey limits. In the near-infrared bands, the initial decline 
in completeness is steeper than Eqn.~(\ref{doubleexp}) can match.
We experimented with fitting a single exponential model and in 
Sect.~\ref{sec:6dfgs} we demonstrate that the luminosity functions are 
unaffected by deviations of this size from the fit.
Individual uncertainties in Fig.~\ref{fig:magcompl} were computed as follows.
Given that completeness is the ratio of  $N_z$ successful redshifts to $N$ 
sources measured, then the uncertainty in completeness per bin is 
\begin{equation}
  \Delta C =  \frac{\sqrt{N(N_z+2)(N-N_z+1)}}{N(N+3)} .
\label{complerror}
\end{equation}

\begin{table}
\begin{center}
\caption{
  Magnitude-dependent completeness parameters Parameters describing the
  best analytic fit to the magnitude-dependent completeness in
  Eqn.~(\ref{doubleexp}).
\label{tab:parameters}
} 
\vspace{6pt}
\begin{tabular}{lcccc}
\hline \hline
Field Completeness & Field   & $\beta_F$ & $\mu_F$ & $\xi_F$ \\
                   & Set $F$ &           & (mags)  & (mags)  \\
\hline
{\bf \kb-band:}   &    &   &   &   \\
$0.9 < R_f \leq 1.0$ &  1  & 0.80  & 14.84  & 12.03  \\
$0.8 < R_f \leq 0.9$ &  2  & 0.74  & 14.72  & 11.96  \\
$0.7 < R_f \leq 0.8$ &  3  & 0.71  & 14.90  & 12.03  \\
$0.6 < R_f \leq 0.7$ &  4  & 0.70  & 15.02  & 12.20  \\
    &   &   &   &    \\
{\bf \hb-band:}   &    &   &   &   \\
 $0.9 <R_f\leq 1.0$ &  1  & 0.85  & 14.66  & 12.27  \\
 $0.8 <R_f\leq 0.9$ &  2  & 0.79  & 14.60  & 12.20  \\
 $0.7 <R_f\leq 0.8$ &  3  & 0.70  & 15.12  & 12.39  \\
 $0.6 <R_f\leq 0.7$ &  4  & 0.87  & 14.24  & 12.37  \\
    &   &   &   &    \\
{\bf \jb-band:}   &    &   &   &   \\
 $0.9 <R_f\leq 1.0$ &  1  & 0.83  & 15.56  & 12.99  \\
 $0.8 <R_f\leq 0.9$ &  2  & 0.78  & 15.32  & 12.97  \\
 $0.7 <R_f\leq 0.8$ &  3  & 0.79  & 15.14  & 13.11  \\
 $0.6 <R_f\leq 0.7$ &  4  & 0.84  & 15.20  & 12.87  \\
    &   &   &   &    \\
{\bf \rf-band:}   &    &   &   &   \\
 $0.9 <R_f\leq 1.0$ &  1  & 0.92  & 16.96  & 15.04  \\
 $0.8 <R_f\leq 0.9$ &  2  & 0.86  & 16.96  & 14.92  \\
 $0.7 <R_f\leq 0.8$ &  3  & 0.83  & 17.25  & 15.16  \\
 $0.6 <R_f\leq 0.7$ &  4  & 0.71  & 20.00  & 14.92  \\
     &   &   &   &    \\
{\bf \bj-band:}   &    &   &   &   \\
 $0.9 <R_f\leq 1.0$ &  1  & 0.87  & 18.29  & 16.10  \\
 $0.8 <R_f\leq 0.9$ &  2  & 0.81  & 18.47  & 16.22  \\
 $0.7 <R_f\leq 0.8$ &  3  & 0.78  & 18.54  & 16.22  \\
 $0.6 <R_f\leq 0.7$ &  4  & 0.70  & 21.29  & 16.40  \\
\hline
\end{tabular}
\end{center}
\end{table}

With the form of the magnitude-dependent completeness known, we are in a
position to determine total completeness, $T_F(\btheta,m)$, as a
function of both sky position $\btheta$ and magnitude $m$. We assume it
is separable function with the general form
\begin{equation}
  T_F(\btheta,m) = S(\btheta) \cdot C_F(m) ,
\label{selectionfunction}
\end{equation}
for all galaxies in a given field group $F$. $S(\btheta)$ is a constant
scaling the completeness of the individual field to that of the total
completeness in the same part of sky,
\begin{equation}
  S(\btheta) = R(\btheta) \cdot
  \frac{\int_{m_{\rm bright}}^{m_{{\rm faint}}} \overline{N(m)} \,\, dm}
  {\int_{m_{\rm bright}}^{m_{{\rm faint}}} C_F(m) \cdot \overline{N(m)}\,\,dm}.
\label{totalcompl}
\end{equation}

\begin{figure*}
\plotfull{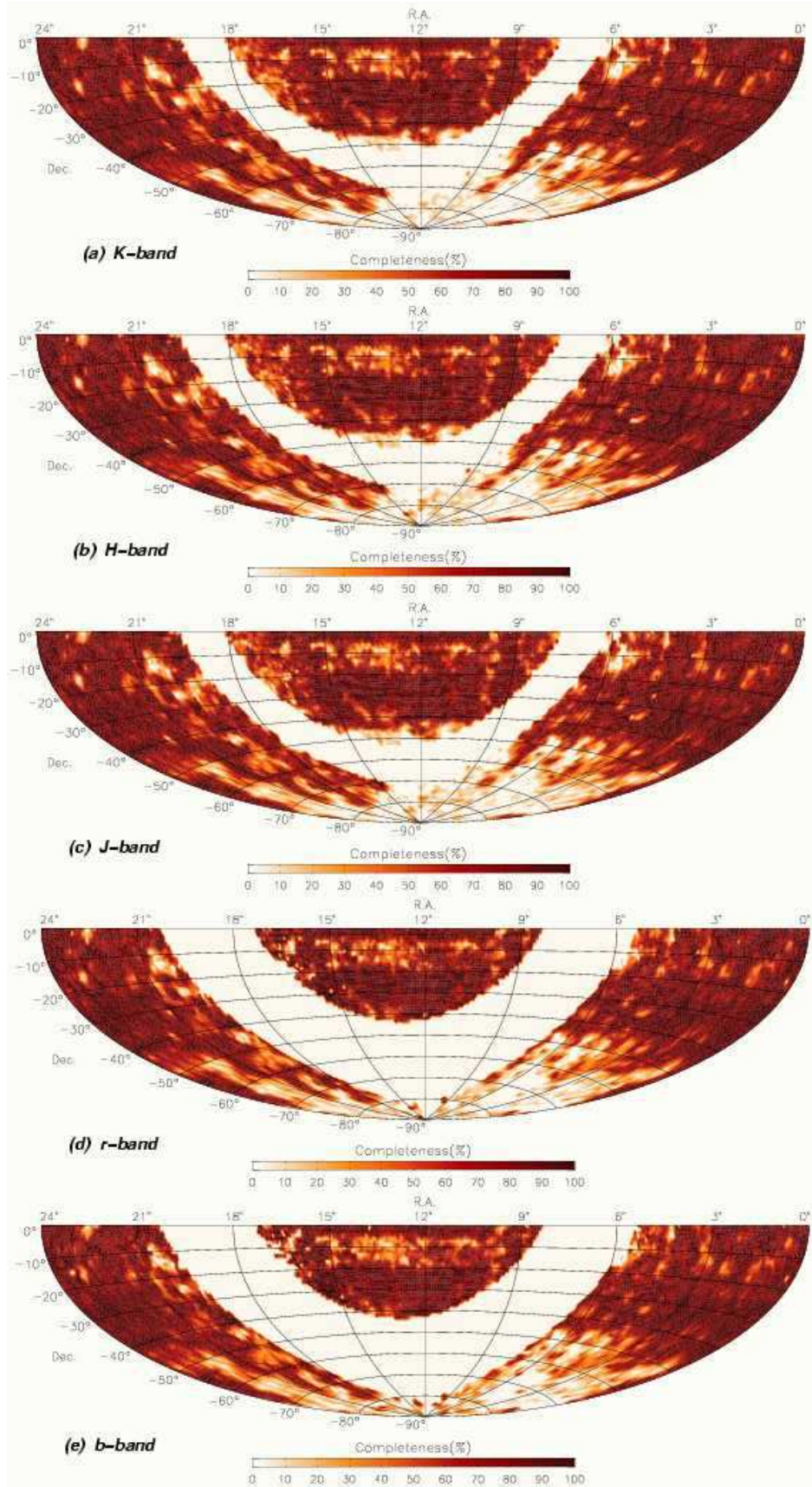}{0.7}
\caption{
  Actual survey completeness on the sky, $R(\btheta)$, for each of the
  \khjrb\ samples used in this paper.}
\label{fig:skycompl}
\end{figure*}

Actual sky completeness, $R(\btheta)$, is the ratio of extragalactic
redshifts ($cz \geq 750\kms$) at sky position $\btheta$, to the number
of confirmed or potentially extragalactic sources at the same position.
It is measured in exactly the same manner as field completeness $R(f)$,
except on 
a uniform $1.5^\circ$ grid of circular regions of diameter $2^\circ$.
This geometry was unrelated to the placement of actual 6dFGS fields or the way in
which they overlapped. Figure \ref{fig:skycompl} shows
the redshift completeness of the all \khjrb\ samples. Typically there is
little variation between bands.

$\overline{N(m)}$ is the mean number of galaxies per magnitude per field
area, which we derive in Section~\ref{sec:numcounts}. The magnitude
limits of the survey are given by $m_{{\rm bright}}$ and $m_{{\rm faint}}$.

In the case of the \vmax\ luminosity estimator \citep{schmidt68}, the
inverse of the total completeness, $(T_F(\btheta_i,m_i))^{-1}$, is used
to weight the inverse maximum accessible volume of each galaxy $i$. For
the STY \cite{sandage79} and SWML \citep{efstathiou88} estimators,
$T_F(\btheta,m)$ is re-expressed in terms of absolute magnitude $M$ for
each galaxy using
\begin{equation}
  M = m -5 \log [d_L(z_i)] - 25 - k(z_i) .
\label{magdiff}  
\end{equation}
Here, $d_L(z_i)$ is the luminosity distance in Mpc, $k(z_i)$ is the
$k$-correction and $z_i$ the redshift of the $i$th galaxy. Equation~(\ref{magdiff})
is written in a form that ignores the evolutionary corrections that become significant at
higher redshift. $T_F(\btheta,M)$ then weights the number density for all $M$.


\section{Luminosity Function Preliminaries}
\label{sec:lfderivation}

\subsection{Number Counts and Normalisation}
\label{sec:numcounts}

Large scale structure plays an important role in defining the overall
normalisation of any LF, and is a crucial factor in determining the
optimal depth and sky coverage of the survey. Arguably it is the single
biggest contributor to discrepancies between recent LFs and related
quantities \citep[see e.g.][]{cole01,norberg02,bell03,driver04,driver05}. While
recent luminosity function determinations have been able to exploit
surveys of unprecedented depth and coverage, cosmic structure is always
an issue for the faintest galaxies of any survey. \citeauthor{cole01}
estimate that cosmic variance could introduce systematic errors as large
as 15 percent in number counts from their sample combining 2MASS NIR
photometry with 2dFGRS redshifts. \cite{bell03} has found the SDSS Early
Data Release to be 8\,percent overdense with respect to the 2MASS
extended source counts over the whole sky, and the 2dFGRS to be slightly
under-dense.

To date, the 6dFGS has covered about a quarter of the sky, or $\sim 2$
to 3~sr depending on passband. Compared to recent near-infrared
(\kb-band) samples, the 6dFGS currently covers about $\sim 15$ to 20
times the area of \citet{cole01} and \citet{bell03} and around
30~percent more that of \citet{kochanek01}. In terms of sample size, the
6dFGS is $\sim 10$---15 times larger than \citet{kochanek01} and
\citet{bell03} and a factor of 4 greater than \citet{cole01}. At optical
magnitudes, current 6dFGS sky coverage quadruples that used for the
2dFGRS \citet{norberg02} and SDSS \citet{blanton03} luminosity
functions, although sample size is roughly half. When complete, the sky
coverage of 6dFGS will be almost double its present extent.

\begin{figure}
\plotone{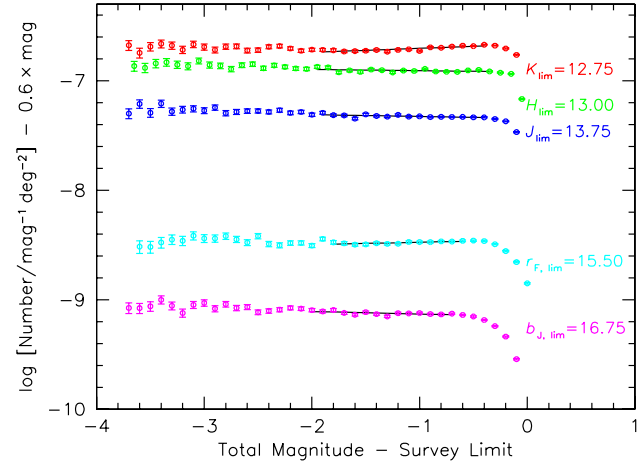}
\caption{
  Differential number counts for the \khjrb\ samples of the 6dFGS.  The
  Euclidean slope of 0.6 has been subtracted for emphasis.  Magnitudes
  have been corrected for Galactic extinction and are expressed as an
  offset from each survey limit (indicated). The solid lines show the
  straight lines of best fit given in Table~\ref{tab:numcounts}. Poisson
  errors are shown.}
\label{fig:numbercounts}
\end{figure}

Figure~\ref{fig:numbercounts} shows the differential number counts for
6dFGS sources after correction for Galactic extinction (Sect.~\ref{extin}).  
The roll-over near the survey limit is the result of the extinction corrections
blurring the sharp survey cut-offs imposed on the original uncorrected
magnitudes. Unlike imaging number count surveys, we have been able to
use our redshift information to remove stellar contamination from the
counts. The counts were averaged over the effective area of the 6dFGS,
obtained by scaling the eventual survey area by the fraction of targets
in hand. Table~\ref{tab:numcounts} gives our linear least-squares fit
parameters $a,b$ for the form
\begin{equation}
  N(m) = 10^{a m + b}
\end{equation}
where $m$ is magnitude and $N(m)$ is the differential surface density
(\magdeg). The counts are flat in all bands with the exception of \kb,
(and to a lesser extent, \rf). This rise has also been seen in the 2MASS
number counts published previously \citep[e.g.][]{cole01,jarrett04}, and
was initially thought to be due to the well-known upwards bias of number
count data in regimes of declining signal-to-noise \citep{murdoch73}. We
tested this by generating artificial random samples with the same size,
flux and error distribution of the real \kb-band sample. This
Monte-Carlo approach revealed that even uncertainties as large as $\dkb
\sim 0.2$ were unable to raise a nominal Euclidean 0.6 slope by more
than 1\,percent. Typical 2MASS flux errors are closer to $\dkb \sim 0.1$
at $\kb = 12.5$. It is possible that the rise in \kb\ reflects the fact
that the fainter bins are capturing more of the intrinsically lower
surface brightness late-type galaxies, although no such rise is seen in
the higher signal-to-noise $H$ or $J$ counts.

Our source density of 3.2~galaxies\magdeg\ at $\kb = 12.5$ is comparable
to the 3\magdeg\ found by \citet{bell03} and \citet{cole01} for the
2MASS XSC in Kron magnitudes. These are typically 0.1 to 0.2\,mag
brighter than the {\sl original} total magnitudes of 2MASS
\citep{cole01}, although it should be noted that the total magnitudes we
now use for 6dFGS are the recently revised ones (Jarrett, priv.\ comm.).
The 6dFGS number counts are identical to 2MASS number counts at $\kiso =
12.2$ \citep{jarrett04}, allowing for the fact that 2MASS isophotal
magnitudes are typically 0.1 to 0.2\,mags fainter than total magnitudes.
Like \citeauthor{bell03}, we find that the Sloan Early Data Release
counts are somewhat higher than average \citep[3.5\magdeg;][]{bell03},
the likely result of an overdensity in its comparatively smaller survey
region (370\sqdeg). In the optical, the 6dFGS \bj-band counts of
2.9\magdeg\ ($\bj=16.0$) are closer to the 2dFGRS counts in the
SDSS--2dFGRS overlap region \citep[2.9\magdeg;][]{norberg02}, although
the NGP and SGP 2dFGRS regions (3.3\magdeg) remain within the errors.
Ultimately, such comparisons are limited by the systematics of different
galaxy photometry techniques and magnitude definitions.

\begin{table}
\begin{center}
\caption{
  Best fit parameters to extinction-corrected 6dFGS number count data.
\label{tab:numcounts}
} 
\vspace{6pt}
\begin{tabular}{cccccc}
\hline \hline
Band &  \multicolumn{2}{c}{Fit range} & Number of & \multicolumn{2}{c}{Fit Parameters} \\
     &  (mag)          & (mag)        & sources   &  $a$  &  $b$ \\
\hline
\kb\  & 10.75  & 12.35 &  113988 & 0.636 & -7.132 \\
\hb\  & 10.95  & 12.65 &   90317 & 0.588 & -6.769 \\
\jb\  & 11.75  & 13.40 &   93831 & 0.583 & -7.114 \\
\rf\  & 13.60  & 14.90 &   64043 & 0.621 & -8.782 \\
\bj\  & 14.70  & 16.10 &   66833 & 0.577 & -8.769 \\
\hline
\end{tabular}
\end{center}
\end{table}

\subsection{Disentangling Hubble and Peculiar Motions}

The line-of-sight velocity measured for a galaxy has two components: the
Hubble flow due to the expansion of the universe, and the peculiar
velocity due to the combined gravitational attraction of all large-scale
structures. To infer true luminosity distances from the Hubble flow
velocity, \citep[assuming $z \ll 1$;][]{hubble34}, we must first remove
the peculiar motion component, especially given the low redshifts for
much of our sample.

We have adopted field flow corrections from software written by
J.~P.~Huchra for the HST Key Project to measure $H_0$ \citep[Appendix~A
of][]{mould00}. This uses a simple linear flow field with three
spherical mass concentrations representing Virgo ($cz = 1\,035\kms$),
the Great Attractor (GA; $4\,600\kms$) and the Shapley Supercluster
(SSC; $13\,800\kms$). We also examined the flow models of
\citet{tonry00} and \cite{burstein89}, but did not use them since they
neglect the SSC, which is a significant influence on much of the 6dFGS
volume.

 \begin{figure}
\plotone{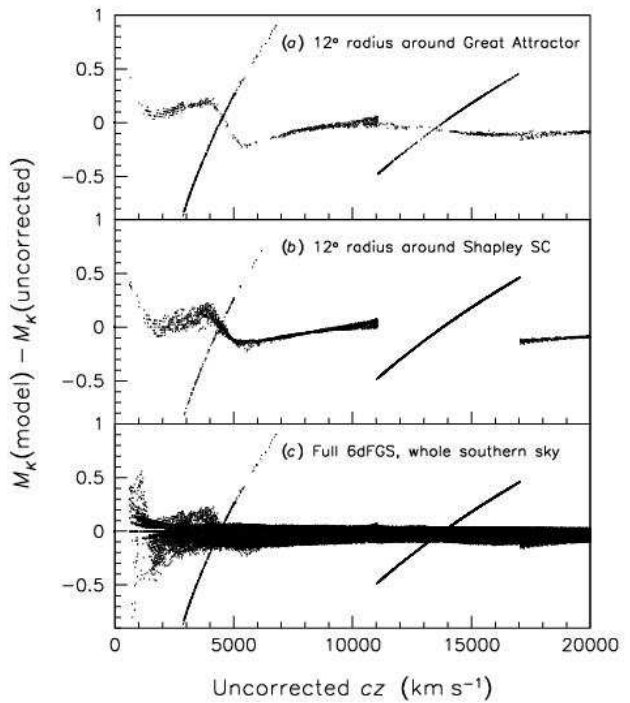}
\caption{
  Change in magnitude resulting from peculiar velocity corrections in
  two key sky directions (the Great Attractor and Shapley Supercluster,
  $a$ and $b$), and for the entire 6dFGS sample ($c$). These show the
  difference between absolute magnitudes derived from the
  flow-model-corrected $cz$ values and uncorrected $cz$ values. The
  diagonal bands are galaxies that the model has identified with the
  Great Attractor ($4\,600$\kms) or Shapley Supercluster ($\sim
  13\,800$\kms) mass concentrations, and ascribed the corresponding
  systemic velocity.}
\label{fig:magdiff}
\end{figure}

Peculiar velocity corrections are most significant for $cz \lesssim
10\,000$\kms. Figure~\ref{fig:magdiff} shows how removal of the peculiar
motion alters the absolute magnitudes for galaxies in the direction of
the GA and SSC.  The Virgo cluster direction is not shown, as it lies
beyond the northern limits of the 6dFGS. In the vicinity of the GA,
galaxy luminosities are affected by as much as $\pm 0.2$~mags while for
the SSC the effect is clearly less ($\lesssim 0.1$~mag). For galaxies
within either mass concentration, the correction can be almost a
magnitude. This is because the flow model sets the galaxy velocity to
that of the mass concentration if the galaxy lies within its sphere of
influence.

With median redshifts in excess of 20\,000\kms, most recent surveys do
not include corrections for peculiar velocities.  An exception is
\citet{kochanek01}, who employed the local flow model of \cite{tonry00}.
\citeauthor{kochanek01} also employ a low-$cz$ cut-off of 2000\kms, so
as to exclude those local galaxies with a significant dependence on flow
corrections, at the expense of 5\,percent of their sample and $\sim
1.5$\,mags at their faint-end limit.

\begin{figure}
\plotone{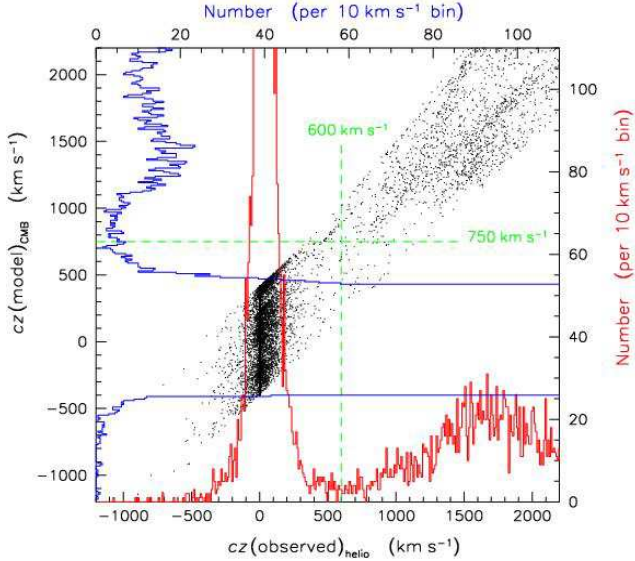}
\caption{
  Effect of flow model on low $cz$ values in relation to the low-$cz$
  cut-off.  The points show observed and model velocities relative to
  the lower and left-hand axes. The histograms show how the distribution
  projects along the same directions relative to the vertical scales
  shown on the right and above. The dashed lines indicate our low-$cz$
  cut-off for both observed (600\kms) and model-corrected $cz$ values
  (750\kms).}
\label{fig:lowvel}
\end{figure}

The nominal 6dFGS low-$cz$ cut-off is 600\kms, set so as to separate the
obvious peak in Galactic sources at $cz \approx 0$\kms\ from the rising
redshift distribution of true extragalactic objects
(Fig.~\ref{fig:lowvel}). However, applying the flow model corrections
broadens the distribution of all velocities, including those sources
that are quite clearly Galactic.  Therefore, we adjust the low-$cz$
limit to 750\kms\ and use this throughout as our minimum acceptable
redshift cut-off. Galaxies above the cut-off are used with their
flow-corrected velocities from the model.

\subsection{Effect of Magnitude Errors}

The 6dFGS has been able to take advantage of the emerging availability
of quality photometry over large areas of the sky. This is particularly
the case for the NIR, where the digital 2MASS photometry of the whole
sky permits a uniformity and precision unprecedented in wide area
redshift surveys. With this kind of precision, we are motivated to
examine whether photometric errors modify the shape of the luminosity
distribution, and if so, by how much. For example, a non-gaussian tail in
the error distribution can influence the bright end of the LF due to its
rapid decline in this region.

\begin{figure}
\plotone{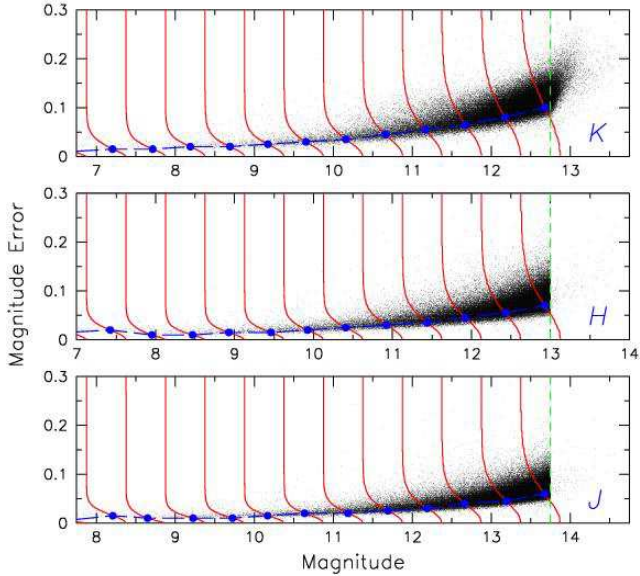}
\caption{
  Distribution of \jhk\ photometry uncertainties for the revised 2MASS
  total magnitudes (points). The large solid blue points and long-dashed 
  blue line both trace the $1\sigma$ deviation of the uncertainties when 
  grouped into 0.5-mag bins. The vertical solid red curves are half-gaussian
  representations of the same $1\sigma$ deviation.
  The vertical short-dashed lines indicate the nominal 6dFGS magnitude limits 
  in each band. There are a small number of galaxies fainter than this
  limit due to the improvements to photometry after target
  selection.} 
\label{fig:magerr}
\end{figure}

\begin{figure}
\plotone{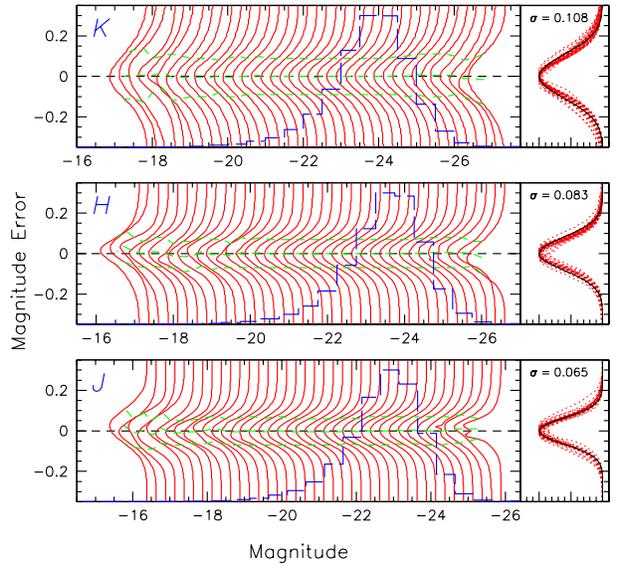}
\caption{
  Distribution of \jhk\ photometric errors as a function of absolute
  total magnitude. The vertical lines are the result of dividing
  galaxies into 0.22-mag bins and co-adding gaussians equal to the
  $1\sigma$ magnitude errors of the relevant galaxies. The short dashed
  lines indicate the $\pm 1\sigma$ deviation and the long dashed
  histogram shows the distribution of 6dFGS sources on the same scale of
  absolute magnitude. The right-hand panels show the superposition of
  magnitude distributions for each bin (dotted lines) with the average
  for all shown by the solid line. The $\sigma$ for this distribution is
  indicated uppermost in each panel.}
\label{fig:abserr}
\end{figure}

Figure~\ref{fig:magerr} shows 2MASS photometric errors for \jhk\ for
6dFGS targets. Typical errors are $\sim 0.1$~mag in \kb\ and $\sim 0.06$
to 0.08~mag in \jb\ and \hb. In terms of absolute magnitude, these
translate fairly uniformly across the range of observed luminosities
(Fig.~\ref{fig:abserr}). Because of this, we apply the mean
photometric error ($\sigma = 0.108, 0.083$ and 0.065 for \khj\ 
respectively) to all luminosities equally when computing its effect on
the LF. Uncertainties in \bj\ and \rf\ are typically $\sim 0.1$,
comparable with \kb.

The functional form most commonly used to match the galaxy luminosity
distribution is that proposed by \citet{schechter76}, which gives the
space density of galaxies per unit absolute magnitude, $\phi(M)$, as
\begin{equation}
  \phi(M)\,dM = 0.4 \ln 10\,\ps\, 
   \frac{ (10^{0.4(M_*-M)})^{\alpha+1} }  {\exp[10^{0.4(M_*-M)}]}\,dM .
\label{schecht}
\end{equation}
The parameter \ps\ is a normalisation constant, while \al\ is the slope
for faint galaxies and \ms\ is a characteristic magnitude at the
transition from the exponential to power law dependence.
While both the Schechter Function and Press-Schechter Mass Function
\citep{press74}
are superficially similar, they are not related in any simple physical way.
The exponential cutoff at the bright end of the Schechter Function is
due to cooling \citep[e.g.][]{rees77} and perhaps feedback by AGN.
If it were caused only by the high mass cutoff of the Press-Schechter
Function then \ms\ would be $\sim 4$ magnitudes brighter.

\begin{figure}
\plotone{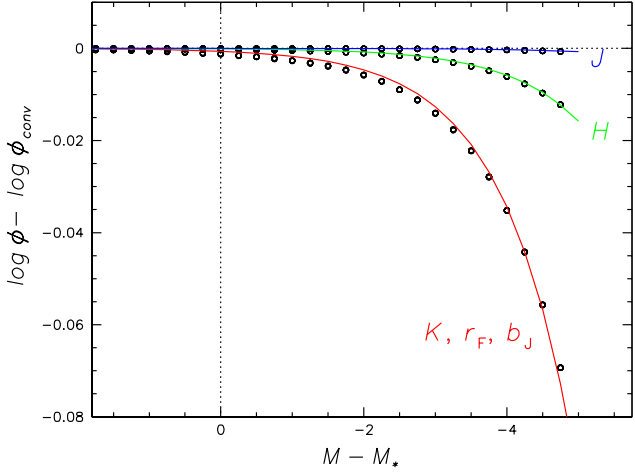}
\caption{
  The open circles show the offset between an ordinary Schechter
  function ($\log \phi$) and one convolved with the gaussian error
  kernels for \khjrb\ ($\log \pc$). The result is independent of \al.
  The solid lines show the functional form of Eqn.~(\ref{schechtconv})
  with $\upsilon = 7.37$ for \kb\rf\bj, 9.15 for \hb, and 12.25 for \jb.
  The vertical dotted line marks the location of \ms.}
\label{fig:conv}
\end{figure}

In order to explore the effect of magnitude errors on this analytic fit, we convolved Schechter
functions across a range of \al\ values with the gaussian error kernels
discussed above. Figure~\ref{fig:conv} shows how a generic Schechter function changes in
$\log \phi$-space when convolved with gaussian magnitude distributions
of $\sigma = 0.07$ to 0.1. As expected, only the brightest luminosities
in the region $M < \ms$ are affected.  Perhaps more surprising is that
the amount of change is {\sl highly insensitive} to the faint-end
slope \al\ for typical values. In the case of \jb, the magnitude errors
are too small to have any appreciable effect on the Schechter fit. In
the interests of computational efficiency we have used the empirical
relationship
\begin{equation}
  \log \pc = \log \phi + e^{M_* - M - \upsilon}
\label{schechtconv}
\end{equation}
with values of $\upsilon = 7.37$ to match the convolution offsets in
\kb\rf\bj, 9.15 for \hb, and 12.25 for \jb. Figure~\ref{fig:conv} shows
each of these curves. These offsets were then applied when finding
optimal Schechter function fits in all bands (Section~\ref{sec:lumfunc}).

\subsection{Extinction and Cosmological Corrections \label{extin}}

One of the primary science drivers for the NIR selection of the main
6dFGS samples was the low impact of Galactic extinction over much of the
sky. We use the maps of \citet{schlegel98} to correct our total
magnitudes using the relative extinction values of $A/A_V = 0.112,
0.176, 0.276, 0.810, 1.236$ for \khjrb\ respectively. Typically this
furnishes extinction corrections of up to $\sim 0.04$\,mag in \kb, $\sim
0.1$\,mag in \jb, and $\sim 0.3$ in \bj\ (recall that the respective
Galactic latitude limits are $|\,b\,| > 10^\circ$ for \jhk\ and $|\,b\,|
> 20^\circ$ for \br). Extinction corrections were applied after the
initial apparent magnitude selection.

Our cosmological $k$-corrections use the values computed by
\citet{poggianti97}, linearly interpolated with redshift. Given the
6dFGS selection bias towards early-types we adopt her $k$-corrections
for an elliptical galaxy with solar metallicity and exponentially
decreasing star-formation rate with $e$-folding time of 1~Gyr. These
corrections are within 0.1\,mag of those used by \citet{kochanek01} for
their \kb-band sample, and also those of \citet{norberg02} for the
2dFGRS \bj-band sample.

We do not apply evolutionary $e$-corrections 
\citep[e.g. ][]{norberg02,blanton03}, since the bulk of our galaxies are
seen at relatively recent cosmic time ($z \lesssim 0.1$).

\section{Luminosity Functions}
\label{sec:lumfunc}

\subsection{6dF Galaxy Survey \label{sec:6dfgs}}

Several techniques have been devised over the years for measuring galaxy
LFs \citep[see e.g.][]{willmer97}. We have adopted the three most
commonly used --- \vmax, STY and SWML --- all of which have merits and
shortcomings. The \vmax\ method \citep{schmidt68,felten76} has the
advantages of simplicity and no {\it a priori} assumption of a
functional form for the luminosity distribution. Unlike the others, it
also yields a fully normalised solution and is insensitive to magnitude
incompleteness. Its main disadvantage is its assumption of a
homogeneous, unclustered galaxy distribution. The STY method
\citep{sandage79} overcomes these difficulties, being a
maximum-likelihood estimator with the added advantage that it does not
require binning. However, it presupposes some explicit functional model
for the luminosity function. The Step-Wise Maximum Likelihood method
\citep[SWML;][]{efstathiou88} has similar advantages to the STY method,
but does not assume an explicit form for the luminosity function. It
produces a binned maximum-likelihood estimate of the LF. STY and SWML
provide shape information but not the overall normalisation in the
manner of \vmax. We refer the interested reader to the individual
references for detailed methodologies.

\begin{figure*}
\plotfull{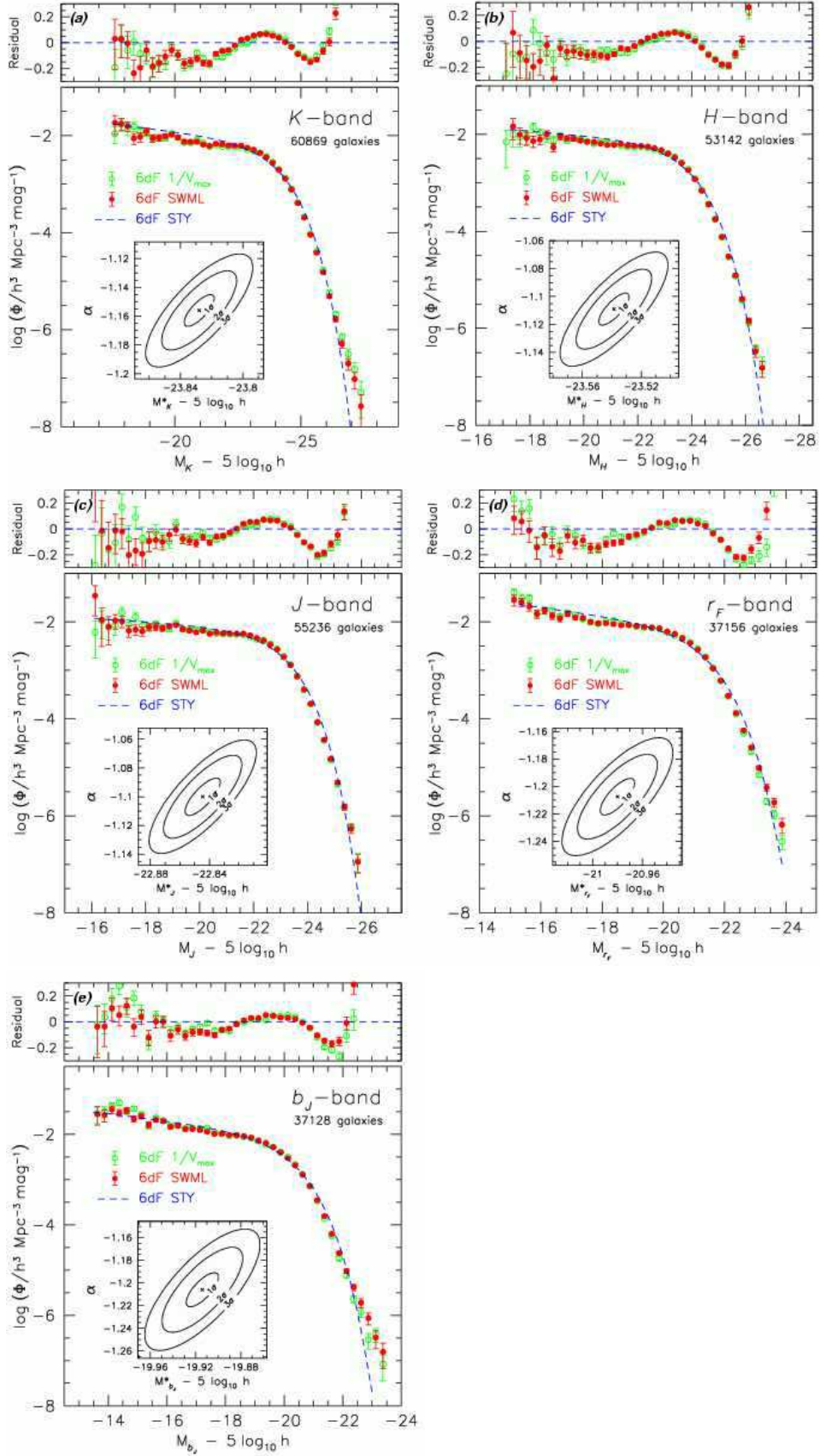}{0.73}
\caption{
  Luminosity functions for the 6dF Galaxy Survey, derived from the
  \vmax\ (green open circles), SWML (red solid circles) and STY methods
  (blue dashed curve). The inset shows the 1, 2 and $3\sigma$ confidence
  contours of the STY fit. The upper panel shows the \vmax\ and SWML
  residuals relative to STY (i.e.\ the deviations from the best-fitting
  Schechter function).}
\label{fig:sty}
\end{figure*}

\begin{table*}
\begin{center}
\caption{
  Schechter function parameters for the STY luminosity functions of 6dFGS. The convolution
  parameter $\upsilon$ used for Eqn.~(\ref{schechtconv}) is given for each band.
\label{tab:schechtfits}
} 
\vspace{3pt}
\begin{tabular}{cc@{\hspace{10mm}}ccc@{\hspace{10mm}}c}
\hline
Sample & Fit Range & $\ms - 5 \log h$ & $\alpha$ & $\log [\ps]$   & $\upsilon$   \\
       & (mag)     & (mag)            &          & ($\invcubicMpc$)     &      \\
\hline
\kb & [-15.50,-28.85] & $-23.83\pm0.03$ & $ -1.16\pm0.04$ & $ -2.126\pm0.005$   &  7.37\\
\hb & [-16.00,-28.50] & $-23.54\pm0.04$ & $ -1.11\pm0.04$ & $ -2.141\pm0.005$   &  9.15\\
\jb & [-15.00,-27.50] & $-22.85\pm0.04$ & $ -1.10\pm0.04$ & $ -2.148\pm0.005$     &  12.25\\
\rf & [-14.00,-25.00] & $-20.98\pm0.05$ & $ -1.21\pm0.04$ & $ -2.081\pm0.006$    & 7.37 \\
\bj & [-12.50,-24.00] & $-19.91\pm0.05$ & $ -1.21\pm0.05$ & $ -1.983\pm0.006$  & 7.37   \\
\hline
\end{tabular}\\
\end{center}
\end{table*}

We have applied the \vmax, STY and SWML methods to each of the 6dFGS
samples in \khjrb.  The STY and SWML LFs were normalised by a $\chi^2$
minimisation with respect to the equivalent \vmax\ distribution.

Sources were excluded if they were outside the redshift range $0.0025 <
z < 0.2$ or had total completeness $T_F(\btheta_i,m_i)<0.6$.
Completeness corrections had much less impact on the \vmax\ results
compared to other methods, a point that has been noted by others
\citep[e.g.][]{kaldare01}. Those sources with extinction-corrected original
magnitudes outside the range $m_{{\rm bright}}$ to $m_{{\rm faint}}$
were also excluded. Here, $m_{{\rm faint}}$ is simply the nominal survey
limit $(12.75, 13.00, 13.75, 15.60, 16.75)$ corrected for extinction in
the direction of the source, while $m_{{\rm bright}} = 8.75$, 9.0, 9.75,
13.0, and 14.0 for \khjrb\ respectively. Particular care was taken to
exclude sources with $\rf \lesssim 13$ or $\bj \lesssim 14$ because of
saturated \rf\ or \bj\ photometry in the original Southern Sky Survey
plate material. The sizes of the samples actually used in the analysis
are given in Table~\ref{tab:targets}.

Figure~\ref{fig:sty} shows the \khjrb\ LFs derived for the 6dFGS via
these methods. Table~\ref{tab:schechtfits} summarises the best-fitting
error-convolved Schechter functions, while Tables~\ref{tab:vmaxswmlKHJ}
and \ref{tab:vmaxswmlRB} list the non-parametric \vmax\ and SWML
distributions. All three methods are in excellent agreement, insofar as
the Schechter functions represent the data. Moving to bluer passbands we
find a slight steepening of the faint-end slope, although not to the
extent that has been reported previously for dense environments. This is
a consequence of the optical samples harbouring a larger population of
late-type galaxies, which are typically of lower luminosity.

The large sample size of the 6dFGS, even at this intermediate stage of
the survey, has allowed us to beat down random uncertainties over a
considerable range of luminosity. As a consequence, we can clearly see
that a Schechter function (Eqn.~\ref{schecht}) is only an approximate
match to the real galaxy luminosity distribution in all bands (upper
panels in Figure~\ref{fig:sty}). Specifically, it is unable to turn over
sharply enough around \ms, and tends to slightly underestimate the true
space density of \ms\ galaxies while at the same time slightly
overestimating that of $M \sim (\ms-1.5)$ and $M \gtrsim (\ms+1.5)$ systems. 
The differences are typically $\sim
15$\,percent around \ms\ and $\sim 30$ to 40\,percent beyond. Similar
outcomes have been found in the LF fitting of deep SDSS data
\citep{bell03,blanton05}, prompting these authors to use double
Schechter functions, or a hybrid Schechter function plus power law fit.

\begin{figure}
\plotone{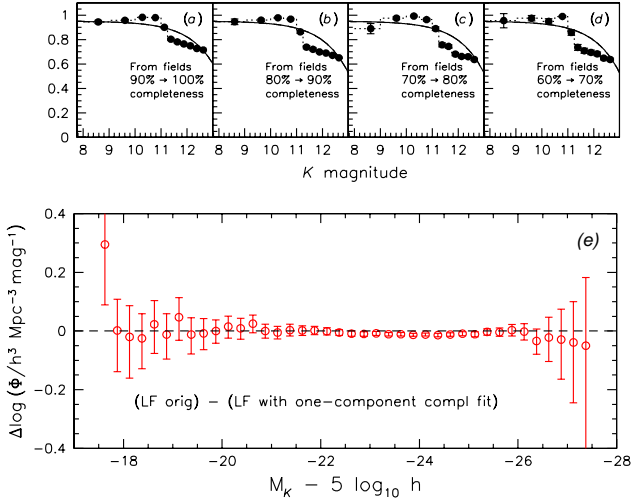}
\caption{
Same \kb-band completeness measurements as shown in ($a$) -- ($d$) 
of Fig.~\ref{fig:magcompl} (top panel), except fit with a single-component exponential.
($e$) The difference between the luminosity function using these
models and that using the double-exponentials of Fig.~\ref{fig:magcompl} for \kb.
The change is negligible in all bins except the very faintest.}
\label{fig:changecompl}
\end{figure}

In Sect.~\ref{sec:selection} we claimed that the luminosity function
is unaffected by small deviations from the model used to fit the magnitude
completeness. Figure~\ref{fig:changecompl} shows the result of
recomputing the \kb-band luminosity function using a single-component
exponential, such as those used by \citet{colless01} for 2dFGRS.
We find there is negligible change in the luminosity function 
(Fig.~\ref{fig:changecompl}$e$), which appears to be fairly insensitive to
the exact form used to model magnitude completeness.

\subsection{Luminosity Function Extremities}

\begin{figure}
\plotone{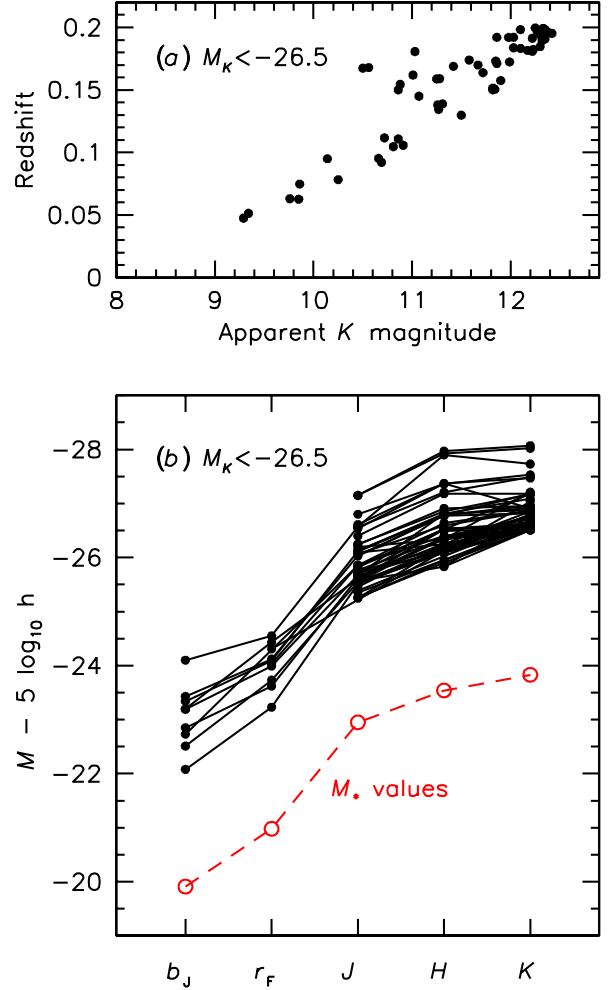}
\caption{($a$) Apparent magnitude-redshift relation for the most
luminous members of the \kb-band selected sample ($\mk < -26.5$).
($b$) Absolute magnitudes in \khjrb\ for the same $\mk < -26.5$
sample. Also shown are the \ms\ values in each band (open circles
and dashed line). Cosmological $k$-corrections have been applied.}
\label{fig:luminous}
\end{figure}

At faint luminosities $M \gtrsim (\ms  + 2)$ we find marginal evidence for
an upturn, as has been suggested by a number of authors for clusters 
\citep[][and references therein]{driver04}. At the bright end we see 
a prominent upturn at $M \lesssim (\ms - 2)$, particularly in the \kb\ and 
\bj\ samples, as has been noted by several authors previously.
 This excess of luminous objects is presumably due to brightest 
cluster galaxies, which are produced by the special merger and accretion 
processes that come into effect in the high density regime at the centre 
of cluster gravitational potentials.  We experimented with fitting this feature 
using the gaussian LF of \citet[][their Eqn.~6.1]{saunders90}, but it only 
improved the bright-end fit at the expense of poorer fitting elsewhere.

We studied the luminous galaxies responsible for the bright-end
upturn in detail. Incorrect redshifts or the presence of a non-gaussian 
tail on the magnitude error distribution could also cause an upturn
if present in our data. Figure \ref{fig:luminous}($a$) shows 
the magnitude-redshift relation for the most luminous galaxies with
$\mk < -26.5$. It demonstrates that they span a range of
of redshift and brightness, as expected for a luminous sample that
should be visible over a large survey volume.
Figure \ref{fig:luminous}($b$) shows the $k$-corrected absolute 
magnitudes in \khjrb\ for the same sample of galaxies. Some galaxies
do not have matching magnitudes in the shorter wavelength bands 
due to non-selection by the original apparent magnitude cut. 
Also shown in Fig.~\ref{fig:luminous}($b$) is the location of \ms\ in 
each band. From this we observe that it is the same luminous 
galaxies that are responsible for the bright-end upturn in all bands.
Hence, galaxy membership of the luminous bins is real and
not the result of an extended tail in the magnitude error distribution.

\begin{figure*}
\plotfull{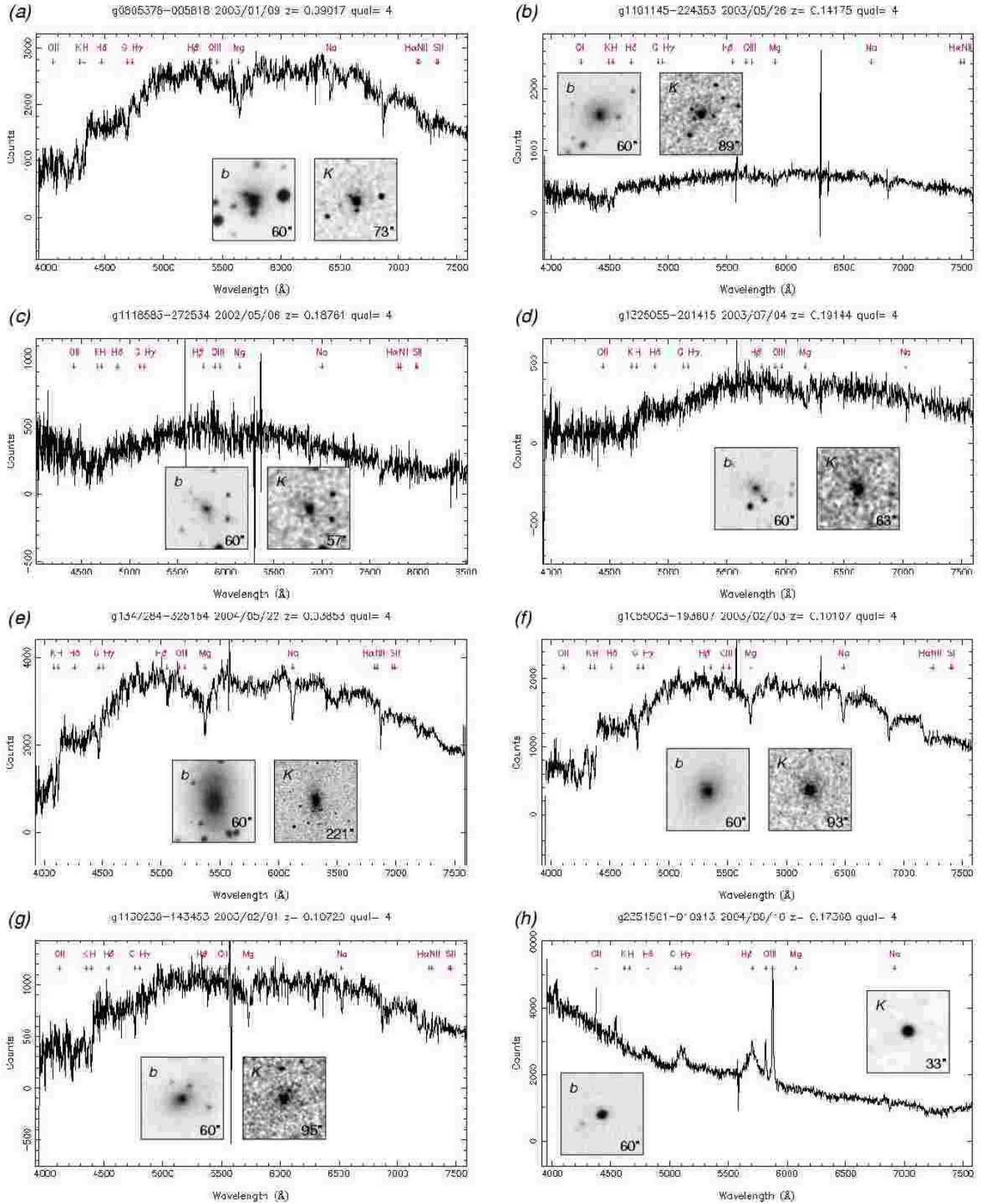}{1.0}
\caption{Spectra of some luminous upturn sources at
$\mk < -26.5$. The labeling of spectral features indicates 
position only and does not imply detection of the feature.
Also shown are postage stamp images in \bj\ and \kb,
taken from SuperCOSMOS and 2MASS respectively. North is up, 
east is left, and image size is indicated in the lower
right-hand corner. 
}
\label{fig:bright}
\end{figure*}

We also paid special attention to the spectra and redshifts of this
luminous sample. Figure \ref{fig:bright} shows a sub-sample of these 
with spectra and \bj\kb\  imaging obtained from
the 6dFGS database. Generally, the spectra have signal-to-noise ratios
that vary from excellent to poor. This is not surprising given the large range
of apparent magnitude seen in Fig.~\ref{fig:luminous}($a$). 
The imaging shows that there are a number of galaxy pairs in this
luminous sample. While they are not close enough to have inflated magnitudes
due to image mergers --- 2MASS can individually identify sources 5\,\arcsec\ 
apart --- it supports the idea that many of these galaxies inhabit dense environments.
In terms of redshift quality, $Q$ \citep{jones04}, 
around half the sample have $Q=4$ indicating
that the redshift measurements are secure. Of the remainder, half
have $Q=3$ while the other redshifts come from the literature.
We can therefore rule out spurious redshifts as a cause of the
bright-end upturn.

\begin{table*}
\begin{center}
\caption{
  Near-infrared \vmax\ and SWML luminosity functions from the 6dFGS.
\label{tab:vmaxswmlKHJ}} 
{\small
\begin{spacing}{1.55}
\vspace{-7mm}
\begin{tabular}{ccc@{\hspace{3mm}}ccc@{\hspace{3mm}}ccc  }
\hline
\multicolumn{3}{c}{{\bf \kb-band}} & \multicolumn{3}{c}{{\bf \hb-band}} & \multicolumn{3}{c}{{\bf \jb-band}} \\ 
\vspace{-1.5mm}
$\mk -5\log h$& \lvmax\ & \lswml\ & $\mh -5\log h$ & \lvmax\ & \lswml\ & $\mj -5\log h$ & \lvmax\ & \lswml\  \\
(mag) & $(\invcubicMpc)$ & $(\invcubicMpc)$ & (mag) & $(\invcubicMpc)$ & $(\invcubicMpc)$ & (mag) & $(\invcubicMpc)$ & $(\invcubicMpc)$ \\
\hline
$-27.375$& $-7.295\,^{+0.232}_{-0.533}$&        $-7.582\,^{+0.233}_{-0.535}$&   $-26.625$& $-6.728\,^{+0.131}_{-0.189}$&        $-6.818\,^{+0.134}_{-0.194}$&   $-25.875$& $-6.932\,^{+0.149}_{-0.228}$&        $-6.948\,^{+0.151}_{-0.232}$\\
$-27.125$& $-6.818\,^{+0.139}_{-0.206}$&        $-7.022\,^{+0.140}_{-0.208}$&   $-26.375$& $-6.432\,^{+0.090}_{-0.113}$&        $-6.474\,^{+0.098}_{-0.127}$&   $-25.625$& $-6.233\,^{+0.074}_{-0.089}$&        $-6.262\,^{+0.080}_{-0.099}$\\
$-26.875$& $-6.504\,^{+0.103}_{-0.135}$&        $-6.701\,^{+0.104}_{-0.137}$&   $-26.125$& $-5.875\,^{+0.052}_{-0.060}$&        $-5.845\,^{+0.056}_{-0.064}$&   $-25.375$& $-5.823\,^{+0.050}_{-0.057}$&        $-5.814\,^{+0.055}_{-0.063}$\\
$-26.625$& $-6.149\,^{+0.069}_{-0.082}$&        $-6.297\,^{+0.071}_{-0.084}$&   $-25.875$& $-5.421\,^{+0.035}_{-0.039}$&        $-5.394\,^{+0.038}_{-0.042}$&   $-25.125$& $-5.354\,^{+0.033}_{-0.036}$&        $-5.313\,^{+0.036}_{-0.039}$\\
$-26.375$& $-5.680\,^{+0.041}_{-0.045}$&        $-5.785\,^{+0.044}_{-0.049}$&   $-25.625$& $-4.922\,^{+0.024}_{-0.025}$&        $-4.910\,^{+0.026}_{-0.028}$&   $-24.875$& $-4.853\,^{+0.022}_{-0.023}$&        $-4.822\,^{+0.024}_{-0.025}$\\
$-26.125$& $-5.233\,^{+0.027}_{-0.029}$&        $-5.313\,^{+0.030}_{-0.032}$&   $-25.375$& $-4.528\,^{+0.018}_{-0.018}$&        $-4.520\,^{+0.019}_{-0.020}$&   $-24.625$& $-4.444\,^{+0.016}_{-0.017}$&        $-4.432\,^{+0.018}_{-0.019}$\\
$-25.875$& $-4.770\,^{+0.019}_{-0.020}$&        $-4.816\,^{+0.021}_{-0.022}$&   $-25.125$& $-4.124\,^{+0.013}_{-0.013}$&        $-4.115\,^{+0.014}_{-0.015}$&   $-24.375$& $-4.080\,^{+0.012}_{-0.013}$&        $-4.066\,^{+0.013}_{-0.014}$\\
$-25.625$& $-4.383\,^{+0.014}_{-0.014}$&        $-4.412\,^{+0.016}_{-0.016}$&   $-24.875$& $-3.747\,^{+0.010}_{-0.010}$&        $-3.752\,^{+0.011}_{-0.011}$&   $-24.125$& $-3.695\,^{+0.009}_{-0.010}$&        $-3.687\,^{+0.010}_{-0.010}$\\
$-25.375$& $-4.027\,^{+0.011}_{-0.011}$&        $-4.043\,^{+0.012}_{-0.013}$&   $-24.625$& $-3.437\,^{+0.008}_{-0.008}$&        $-3.443\,^{+0.009}_{-0.009}$&   $-23.875$& $-3.396\,^{+0.008}_{-0.008}$&        $-3.396\,^{+0.008}_{-0.009}$\\
$-25.125$& $-3.680\,^{+0.009}_{-0.009}$&        $-3.689\,^{+0.010}_{-0.010}$&   $-24.375$& $-3.147\,^{+0.007}_{-0.007}$&        $-3.163\,^{+0.007}_{-0.008}$&   $-23.625$& $-3.115\,^{+0.007}_{-0.007}$&        $-3.125\,^{+0.007}_{-0.007}$\\
$-24.875$& $-3.382\,^{+0.007}_{-0.007}$&        $-3.390\,^{+0.008}_{-0.008}$&   $-24.125$& $-2.909\,^{+0.006}_{-0.006}$&        $-2.925\,^{+0.007}_{-0.007}$&   $-23.375$& $-2.878\,^{+0.006}_{-0.006}$&        $-2.890\,^{+0.006}_{-0.007}$\\
$-24.625$& $-3.100\,^{+0.006}_{-0.006}$&        $-3.113\,^{+0.007}_{-0.007}$&   $-23.875$& $-2.721\,^{+0.006}_{-0.006}$&        $-2.737\,^{+0.006}_{-0.006}$&   $-23.125$& $-2.712\,^{+0.006}_{-0.006}$&        $-2.722\,^{+0.006}_{-0.006}$\\
$-24.375$& $-2.871\,^{+0.006}_{-0.006}$&        $-2.889\,^{+0.006}_{-0.006}$&   $-23.625$& $-2.582\,^{+0.006}_{-0.006}$&        $-2.592\,^{+0.006}_{-0.006}$&   $-22.875$& $-2.558\,^{+0.006}_{-0.006}$&        $-2.569\,^{+0.006}_{-0.006}$\\
$-24.125$& $-2.696\,^{+0.005}_{-0.005}$&        $-2.704\,^{+0.006}_{-0.006}$&   $-23.375$& $-2.469\,^{+0.006}_{-0.006}$&        $-2.475\,^{+0.006}_{-0.006}$&   $-22.625$& $-2.455\,^{+0.006}_{-0.006}$&        $-2.468\,^{+0.006}_{-0.006}$\\
$-23.875$& $-2.547\,^{+0.005}_{-0.005}$&        $-2.561\,^{+0.006}_{-0.006}$&   $-23.125$& $-2.401\,^{+0.006}_{-0.006}$&        $-2.395\,^{+0.007}_{-0.007}$&   $-22.375$& $-2.381\,^{+0.006}_{-0.006}$&        $-2.380\,^{+0.007}_{-0.007}$\\
$-23.625$& $-2.435\,^{+0.006}_{-0.006}$&        $-2.442\,^{+0.006}_{-0.006}$&   $-22.875$& $-2.358\,^{+0.007}_{-0.007}$&        $-2.327\,^{+0.008}_{-0.008}$&   $-22.125$& $-2.356\,^{+0.007}_{-0.007}$&        $-2.331\,^{+0.008}_{-0.008}$\\
$-23.375$& $-2.357\,^{+0.006}_{-0.006}$&        $-2.353\,^{+0.006}_{-0.007}$&   $-22.625$& $-2.314\,^{+0.008}_{-0.008}$&        $-2.286\,^{+0.009}_{-0.009}$&   $-21.875$& $-2.306\,^{+0.008}_{-0.008}$&        $-2.275\,^{+0.008}_{-0.009}$\\
$-23.125$& $-2.314\,^{+0.007}_{-0.007}$&        $-2.291\,^{+0.007}_{-0.007}$&   $-22.375$& $-2.274\,^{+0.009}_{-0.009}$&        $-2.253\,^{+0.009}_{-0.010}$&   $-21.625$& $-2.261\,^{+0.009}_{-0.009}$&        $-2.243\,^{+0.009}_{-0.010}$\\
$-22.875$& $-2.292\,^{+0.008}_{-0.008}$&        $-2.251\,^{+0.008}_{-0.008}$&   $-22.125$& $-2.254\,^{+0.010}_{-0.011}$&        $-2.240\,^{+0.011}_{-0.011}$&   $-21.375$& $-2.256\,^{+0.010}_{-0.011}$&        $-2.242\,^{+0.011}_{-0.011}$\\
$-22.625$& $-2.240\,^{+0.009}_{-0.009}$&        $-2.207\,^{+0.009}_{-0.009}$&   $-21.875$& $-2.262\,^{+0.012}_{-0.013}$&        $-2.236\,^{+0.013}_{-0.013}$&   $-21.125$& $-2.249\,^{+0.012}_{-0.013}$&        $-2.237\,^{+0.013}_{-0.013}$\\
$-22.375$& $-2.230\,^{+0.010}_{-0.010}$&        $-2.205\,^{+0.011}_{-0.011}$&   $-21.625$& $-2.258\,^{+0.014}_{-0.015}$&        $-2.230\,^{+0.015}_{-0.015}$&   $-20.875$& $-2.245\,^{+0.014}_{-0.015}$&        $-2.235\,^{+0.015}_{-0.015}$\\
$-22.125$& $-2.222\,^{+0.012}_{-0.012}$&        $-2.203\,^{+0.012}_{-0.013}$&   $-21.375$& $-2.232\,^{+0.016}_{-0.017}$&        $-2.196\,^{+0.017}_{-0.017}$&   $-20.625$& $-2.239\,^{+0.017}_{-0.017}$&        $-2.218\,^{+0.017}_{-0.018}$\\
$-21.875$& $-2.222\,^{+0.014}_{-0.014}$&        $-2.173\,^{+0.014}_{-0.015}$&   $-21.125$& $-2.198\,^{+0.019}_{-0.019}$&        $-2.219\,^{+0.019}_{-0.020}$&   $-20.375$& $-2.205\,^{+0.019}_{-0.020}$&        $-2.242\,^{+0.020}_{-0.021}$\\
$-21.625$& $-2.189\,^{+0.016}_{-0.017}$&        $-2.166\,^{+0.017}_{-0.017}$&   $-20.875$& $-2.164\,^{+0.021}_{-0.022}$&        $-2.218\,^{+0.022}_{-0.023}$&   $-20.125$& $-2.146\,^{+0.021}_{-0.022}$&        $-2.181\,^{+0.021}_{-0.022}$\\
$-21.375$& $-2.222\,^{+0.020}_{-0.020}$&        $-2.210\,^{+0.020}_{-0.021}$&   $-20.625$& $-2.151\,^{+0.024}_{-0.026}$&        $-2.191\,^{+0.025}_{-0.027}$&   $-19.875$& $-2.157\,^{+0.025}_{-0.026}$&        $-2.212\,^{+0.026}_{-0.027}$\\
$-21.125$& $-2.147\,^{+0.021}_{-0.022}$&        $-2.182\,^{+0.022}_{-0.023}$&   $-20.375$& $-2.192\,^{+0.030}_{-0.033}$&        $-2.158\,^{+0.030}_{-0.032}$&   $-19.625$& $-2.159\,^{+0.029}_{-0.031}$&        $-2.178\,^{+0.030}_{-0.032}$\\
$-20.875$& $-2.084\,^{+0.023}_{-0.024}$&        $-2.126\,^{+0.024}_{-0.025}$&   $-20.125$& $-2.142\,^{+0.034}_{-0.036}$&        $-2.154\,^{+0.035}_{-0.038}$&   $-19.375$& $-2.150\,^{+0.034}_{-0.037}$&        $-2.147\,^{+0.034}_{-0.037}$\\
$-20.625$& $-2.143\,^{+0.029}_{-0.031}$&        $-2.126\,^{+0.029}_{-0.032}$&   $-19.875$& $-2.096\,^{+0.038}_{-0.041}$&        $-2.126\,^{+0.039}_{-0.043}$&   $-19.125$& $-2.024\,^{+0.035}_{-0.038}$&        $-2.065\,^{+0.036}_{-0.040}$\\
$-20.375$& $-2.139\,^{+0.034}_{-0.037}$&        $-2.120\,^{+0.035}_{-0.038}$&   $-19.625$& $-2.081\,^{+0.043}_{-0.048}$&        $-2.093\,^{+0.044}_{-0.050}$&   $-18.875$& $-2.143\,^{+0.046}_{-0.051}$&        $-2.090\,^{+0.044}_{-0.050}$\\
$-20.125$& $-2.037\,^{+0.036}_{-0.039}$&        $-2.034\,^{+0.037}_{-0.040}$&   $-19.375$& $-2.130\,^{+0.053}_{-0.061}$&        $-2.082\,^{+0.051}_{-0.058}$&   $-18.625$& $-2.115\,^{+0.052}_{-0.058}$&        $-2.133\,^{+0.054}_{-0.062}$\\
$-19.875$& $-1.957\,^{+0.038}_{-0.042}$&        $-1.976\,^{+0.041}_{-0.045}$&   $-19.125$& $-2.027\,^{+0.055}_{-0.063}$&        $-2.044\,^{+0.056}_{-0.065}$&   $-18.375$& $-2.047\,^{+0.057}_{-0.066}$&        $-2.107\,^{+0.060}_{-0.069}$\\
$-19.625$& $-2.056\,^{+0.050}_{-0.056}$&        $-2.010\,^{+0.050}_{-0.057}$&   $-18.875$& $-2.117\,^{+0.072}_{-0.086}$&        $-2.264\,^{+0.082}_{-0.101}$&   $-18.125$& $-2.102\,^{+0.070}_{-0.083}$&        $-2.105\,^{+0.069}_{-0.082}$\\
$-19.375$& $-2.044\,^{+0.057}_{-0.066}$&        $-2.041\,^{+0.059}_{-0.068}$&   $-18.625$& $-2.053\,^{+0.079}_{-0.097}$&        $-1.995\,^{+0.071}_{-0.084}$&   $-17.875$& $-2.071\,^{+0.084}_{-0.104}$&        $-2.192\,^{+0.085}_{-0.106}$\\
$-19.125$& $-2.056\,^{+0.067}_{-0.079}$&        $-2.050\,^{+0.069}_{-0.082}$&   $-18.375$& $-1.970\,^{+0.082}_{-0.102}$&        $-2.106\,^{+0.087}_{-0.109}$&   $-17.625$& $-1.897\,^{+0.076}_{-0.093}$&        $-2.157\,^{+0.091}_{-0.116}$\\
$-18.875$& $-1.932\,^{+0.071}_{-0.084}$&        $-1.908\,^{+0.069}_{-0.082}$&   $-18.125$& $-1.855\,^{+0.086}_{-0.107}$&        $-2.138\,^{+0.101}_{-0.132}$&   $-17.375$& $-2.060\,^{+0.106}_{-0.141}$&        $-2.179\,^{+0.104}_{-0.138}$\\
$-18.625$& $-1.916\,^{+0.081}_{-0.099}$&        $-2.025\,^{+0.086}_{-0.107}$&   $-17.875$& $-2.051\,^{+0.119}_{-0.165}$&        $-2.078\,^{+0.112}_{-0.151}$&   $-17.125$& $-1.801\,^{+0.103}_{-0.135}$&        $-1.989\,^{+0.102}_{-0.133}$\\
$-18.375$& $-1.809\,^{+0.084}_{-0.104}$&        $-2.052\,^{+0.099}_{-0.128}$&   $-17.625$& $-2.055\,^{+0.139}_{-0.206}$&        $-2.008\,^{+0.134}_{-0.194}$&   $-16.875$& $-2.065\,^{+0.139}_{-0.206}$&        $-1.972\,^{+0.127}_{-0.181}$\\
$-18.125$& $-1.858\,^{+0.106}_{-0.141}$&        $-1.804\,^{+0.090}_{-0.114}$&   $-17.375$& $-2.003\,^{+0.161}_{-0.257}$&        $-1.839\,^{+0.164}_{-0.267}$&   $-16.625$& $-2.112\,^{+0.161}_{-0.257}$&        $-2.096\,^{+0.200}_{-0.383}$\\
$-17.875$& $-1.757\,^{+0.106}_{-0.141}$&        $-1.753\,^{+0.109}_{-0.146}$&   $-17.125$& $-2.153\,^{+0.232}_{-0.533}$&        $-2.216\,^{+0.304}_{-0.374}$&   $-16.375$& $-1.974\,^{+0.198}_{-0.374}$&        $-1.954\,^{+0.236}_{-0.556}$\\
$-17.625$& $-1.958\,^{+0.139}_{-0.206}$&        $-1.736\,^{+0.152}_{-0.237}$&                   &                                                       &                                                           &  $-16.125$& $-2.212\,^{+0.232}_{-0.533}$&         $-1.459\,^{+0.208}_{-0.415}$\\
\hline \\
\end{tabular}\\
\end{spacing}
}  
\end{center}
\end{table*}

\begin{table*}
\begin{center}
\caption{
  Optical \vmax\ and SWML luminosity functions from the 6dFGS. (Replaces Table 6 of
Jones et al.\ 2006, MNRAS, {\bf 369}, 25.)
\label{tab:vmaxswmlRB}
} 
{\small
\begin{spacing}{1.55}
\vspace{-2mm}
\begin{tabular}{ccc@{\hspace{12mm}}ccc  }
\hline
\multicolumn{3}{c}{{\bf \rf-band}} & \multicolumn{3}{c}{{\bf \bj-band}} \\ 
\vspace{-1.5mm}
$\rf -5\log h$& \lvmax\ & \lswml\  &   $\bj -5\log h$ & \lvmax\ & \lswml\ \\
 (mag) & $(\invcubicMpc)$ & $(\invcubicMpc)$ & (mag) & $(\invcubicMpc)$ & $(\invcubicMpc)$ \\
\hline
$-23.875$& $-6.531\,^{+0.119}_{-0.165}$& 	$-6.182\,^{+0.128}_{-0.182}$& 	$-23.375$& $-7.079\,^{+0.198}_{-0.374}$& 	$-6.809\,^{+0.196}_{-0.365}$\\
$-23.625$& $-5.968\,^{+0.076}_{-0.093}$& 	$-5.722\,^{+0.082}_{-0.102}$& 	$-23.125$& $-6.418\,^{+0.125}_{-0.176}$& 	$-6.492\,^{+0.155}_{-0.244}$\\
$-23.375$& $-5.701\,^{+0.061}_{-0.071}$& 	$-5.415\,^{+0.062}_{-0.072}$& 	$-22.875$& $-6.542\,^{+0.139}_{-0.206}$& 	$-6.058\,^{+0.114}_{-0.154}$\\
$-23.125$& $-5.153\,^{+0.040}_{-0.044}$& 	$-5.012\,^{+0.043}_{-0.047}$& 	$-22.625$& $-5.926\,^{+0.090}_{-0.113}$& 	$-5.720\,^{+0.085}_{-0.106}$\\
$-22.875$& $-4.678\,^{+0.027}_{-0.029}$& 	$-4.591\,^{+0.029}_{-0.032}$& 	$-22.375$& $-5.645\,^{+0.073}_{-0.088}$& 	$-5.379\,^{+0.063}_{-0.074}$\\
$-22.625$& $-4.292\,^{+0.020}_{-0.021}$& 	$-4.237\,^{+0.022}_{-0.023}$& 	$-22.125$& $-5.121\,^{+0.046}_{-0.051}$& 	$-5.021\,^{+0.046}_{-0.051}$\\
$-22.375$& $-3.898\,^{+0.015}_{-0.015}$& 	$-3.879\,^{+0.016}_{-0.017}$& 	$-21.875$& $-4.739\,^{+0.034}_{-0.037}$& 	$-4.624\,^{+0.032}_{-0.035}$\\
$-22.125$& $-3.536\,^{+0.011}_{-0.012}$& 	$-3.522\,^{+0.013}_{-0.013}$& 	$-21.625$& $-4.245\,^{+0.022}_{-0.024}$& 	$-4.198\,^{+0.022}_{-0.024}$\\
$-21.875$& $-3.210\,^{+0.009}_{-0.009}$& 	$-3.211\,^{+0.010}_{-0.010}$& 	$-21.375$& $-3.856\,^{+0.016}_{-0.017}$& 	$-3.805\,^{+0.016}_{-0.017}$\\
$-21.625$& $-2.944\,^{+0.008}_{-0.008}$& 	$-2.960\,^{+0.009}_{-0.009}$& 	$-21.125$& $-3.472\,^{+0.012}_{-0.012}$& 	$-3.455\,^{+0.013}_{-0.013}$\\
$-21.375$& $-2.718\,^{+0.007}_{-0.007}$& 	$-2.732\,^{+0.008}_{-0.008}$& 	$-20.875$& $-3.144\,^{+0.009}_{-0.010}$& 	$-3.140\,^{+0.010}_{-0.011}$\\
$-21.125$& $-2.552\,^{+0.007}_{-0.007}$& 	$-2.574\,^{+0.008}_{-0.008}$& 	$-20.625$& $-2.884\,^{+0.008}_{-0.008}$& 	$-2.889\,^{+0.009}_{-0.009}$\\
$-20.875$& $-2.425\,^{+0.007}_{-0.007}$& 	$-2.435\,^{+0.008}_{-0.008}$& 	$-20.375$& $-2.667\,^{+0.007}_{-0.007}$& 	$-2.684\,^{+0.008}_{-0.008}$\\
$-20.625$& $-2.333\,^{+0.007}_{-0.007}$& 	$-2.338\,^{+0.008}_{-0.008}$& 	$-20.125$& $-2.506\,^{+0.007}_{-0.007}$& 	$-2.527\,^{+0.008}_{-0.008}$\\
$-20.375$& $-2.280\,^{+0.008}_{-0.008}$& 	$-2.246\,^{+0.009}_{-0.009}$& 	$-19.875$& $-2.391\,^{+0.007}_{-0.007}$& 	$-2.400\,^{+0.008}_{-0.008}$\\
$-20.125$& $-2.216\,^{+0.009}_{-0.009}$& 	$-2.198\,^{+0.009}_{-0.010}$& 	$-19.625$& $-2.281\,^{+0.007}_{-0.007}$& 	$-2.283\,^{+0.008}_{-0.008}$\\
$-19.875$& $-2.140\,^{+0.009}_{-0.010}$& 	$-2.131\,^{+0.010}_{-0.010}$& 	$-19.375$& $-2.229\,^{+0.008}_{-0.008}$& 	$-2.190\,^{+0.009}_{-0.009}$\\
$-19.625$& $-2.120\,^{+0.011}_{-0.011}$& 	$-2.119\,^{+0.011}_{-0.012}$& 	$-19.125$& $-2.148\,^{+0.009}_{-0.009}$& 	$-2.139\,^{+0.009}_{-0.010}$\\
$-19.375$& $-2.109\,^{+0.012}_{-0.013}$& 	$-2.099\,^{+0.013}_{-0.014}$& 	$-18.875$& $-2.087\,^{+0.009}_{-0.009}$& 	$-2.075\,^{+0.010}_{-0.010}$\\
$-19.125$& $-2.093\,^{+0.015}_{-0.015}$& 	$-2.082\,^{+0.015}_{-0.016}$& 	$-18.625$& $-2.055\,^{+0.010}_{-0.011}$& 	$-2.041\,^{+0.011}_{-0.011}$\\
$-18.875$& $-2.067\,^{+0.017}_{-0.017}$& 	$-2.061\,^{+0.017}_{-0.018}$& 	$-18.375$& $-2.019\,^{+0.012}_{-0.012}$& 	$-2.021\,^{+0.013}_{-0.013}$\\
$-18.625$& $-2.033\,^{+0.019}_{-0.019}$& 	$-2.070\,^{+0.020}_{-0.021}$& 	$-18.125$& $-2.030\,^{+0.014}_{-0.014}$& 	$-2.014\,^{+0.015}_{-0.015}$\\
$-18.375$& $-1.972\,^{+0.020}_{-0.021}$& 	$-2.045\,^{+0.022}_{-0.023}$& 	$-17.875$& $-1.989\,^{+0.015}_{-0.016}$& 	$-1.986\,^{+0.016}_{-0.017}$\\
$-18.125$& $-1.989\,^{+0.024}_{-0.025}$& 	$-2.025\,^{+0.025}_{-0.027}$& 	$-17.625$& $-1.958\,^{+0.017}_{-0.018}$& 	$-1.990\,^{+0.018}_{-0.019}$\\
$-17.875$& $-2.038\,^{+0.030}_{-0.032}$& 	$-2.024\,^{+0.031}_{-0.033}$& 	$-17.375$& $-1.867\,^{+0.018}_{-0.019}$& 	$-1.945\,^{+0.020}_{-0.021}$\\
$-17.625$& $-2.006\,^{+0.035}_{-0.038}$& 	$-1.999\,^{+0.036}_{-0.039}$& 	$-17.125$& $-1.872\,^{+0.021}_{-0.022}$& 	$-1.907\,^{+0.022}_{-0.023}$\\
$-17.375$& $-1.940\,^{+0.037}_{-0.040}$& 	$-1.917\,^{+0.038}_{-0.042}$& 	$-16.875$& $-1.861\,^{+0.025}_{-0.026}$& 	$-1.883\,^{+0.026}_{-0.028}$\\
$-17.125$& $-1.858\,^{+0.039}_{-0.043}$& 	$-1.911\,^{+0.043}_{-0.048}$& 	$-16.625$& $-1.866\,^{+0.029}_{-0.031}$& 	$-1.885\,^{+0.031}_{-0.034}$\\
$-16.875$& $-1.805\,^{+0.044}_{-0.049}$& 	$-1.834\,^{+0.046}_{-0.052}$& 	$-16.375$& $-1.796\,^{+0.031}_{-0.034}$& 	$-1.813\,^{+0.033}_{-0.036}$\\
$-16.625$& $-1.884\,^{+0.055}_{-0.063}$& 	$-1.928\,^{+0.057}_{-0.066}$& 	$-16.125$& $-1.802\,^{+0.037}_{-0.040}$& 	$-1.836\,^{+0.039}_{-0.043}$\\
$-16.375$& $-1.777\,^{+0.056}_{-0.065}$& 	$-1.874\,^{+0.062}_{-0.072}$& 	$-15.875$& $-1.700\,^{+0.038}_{-0.042}$& 	$-1.705\,^{+0.039}_{-0.043}$\\
$-16.125$& $-1.766\,^{+0.065}_{-0.076}$& 	$-1.762\,^{+0.063}_{-0.074}$& 	$-15.625$& $-1.640\,^{+0.042}_{-0.046}$& 	$-1.682\,^{+0.044}_{-0.049}$\\
$-15.875$& $-1.830\,^{+0.078}_{-0.095}$& 	$-1.834\,^{+0.076}_{-0.092}$& 	$-15.375$& $-1.817\,^{+0.056}_{-0.064}$& 	$-1.785\,^{+0.055}_{-0.063}$\\
$-15.625$& $-1.499\,^{+0.068}_{-0.080}$& 	$-1.682\,^{+0.075}_{-0.091}$& 	$-15.125$& $-1.547\,^{+0.050}_{-0.057}$& 	$-1.600\,^{+0.052}_{-0.059}$\\
$-15.375$& $-1.511\,^{+0.076}_{-0.093}$& 	$-1.590\,^{+0.080}_{-0.099}$& 	$-14.875$& $-1.438\,^{+0.051}_{-0.058}$& 	$-1.658\,^{+0.062}_{-0.072}$\\
$-15.125$& $-1.395\,^{+0.081}_{-0.099}$& 	$-1.544\,^{+0.096}_{-0.124}$& 	$-14.625$& $-1.490\,^{+0.064}_{-0.075}$& 	$-1.474\,^{+0.058}_{-0.067}$\\
&     &      &                                                                  $-14.375$& $-1.303\,^{+0.056}_{-0.065}$& 	$-1.526\,^{+0.070}_{-0.083}$\\
&     &      &	                                                                $-14.125$& $-1.380\,^{+0.081}_{-0.099}$& 	$-1.452\,^{+0.078}_{-0.096}$\\
&     &      &                                                               	$-13.875$& $-1.502\,^{+0.100}_{-0.130}$& 	$-1.573\,^{+0.114}_{-0.156}$\\
&     &      &                                                              	$-13.625$& $-1.536\,^{+0.149}_{-0.228}$& 	$-1.554\,^{+0.154}_{-0.242}$\\
\hline \\
\end{tabular}\\
\end{spacing}
}  
\end{center}
\end{table*}

\subsection{Comparison to Other Surveys}

\begin{figure*}
\plotfull{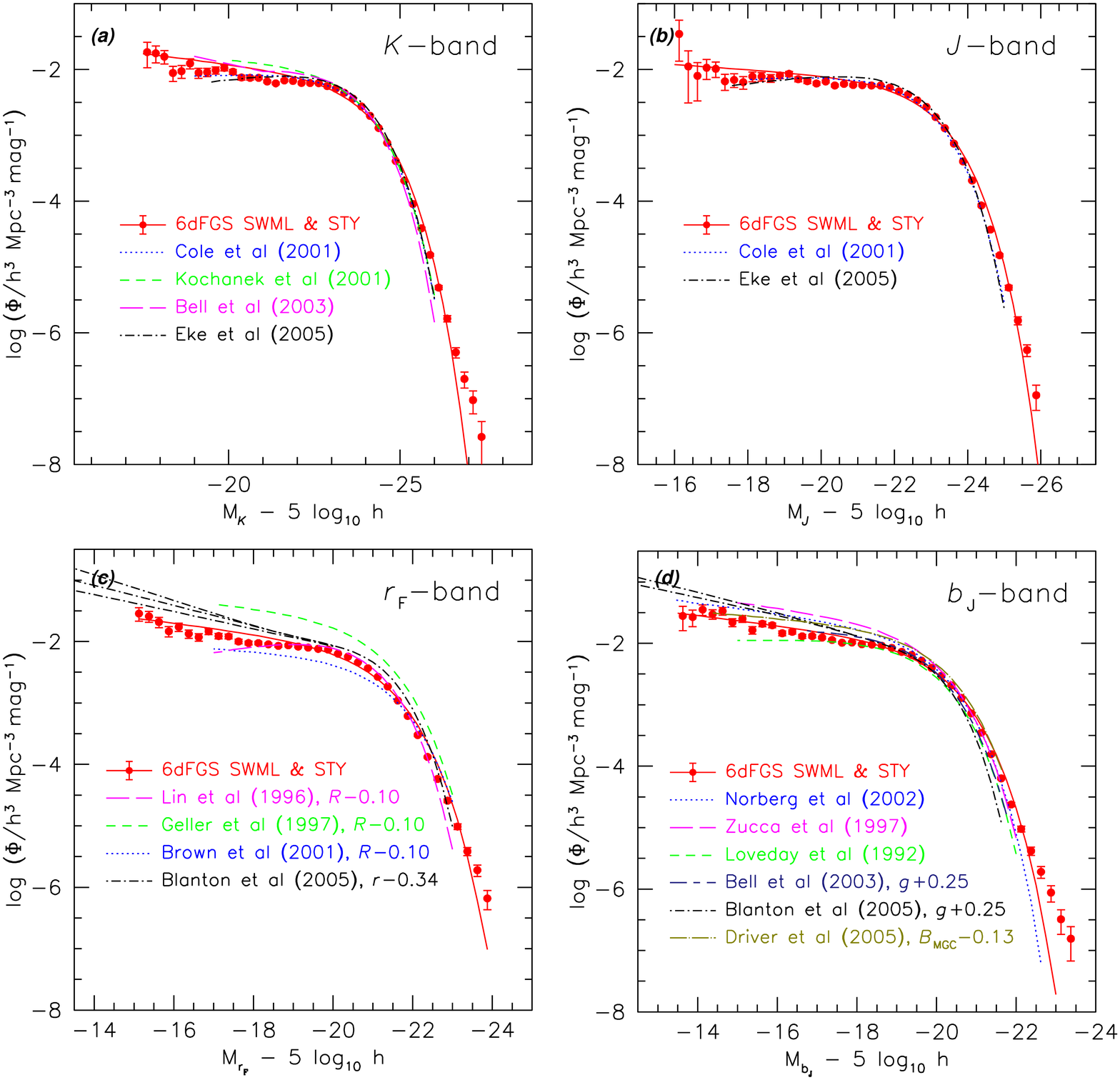}{0.90}
\caption{
  Comparison between the 6dFGS luminosity functions for \kb\jb\rf\bj\ 
  and those of other surveys. Only the STY and SWML fits to 6dFGS have
  been reproduced.  Ordinary Schechter functions are shown throughout
  except for the double Schechter function of \citeauthor{blanton05}
  ($c$ and $d$) and the composite Schechter-and-power-law form of
  \citeauthor{bell03} for \kb\ ($a$).  The \kb and \jb\ Kron magnitudes
  of \citeauthor{cole01} have been converted to the total magnitudes
  used here by subtracting $0.135$\,mag; the isophotal magnitudes of
  \citeauthor{kochanek01} have similarly been corrected by subtracting
  $0.185$\,mag. The passband transformations $\rf = r-0.34$ (for
  \citeauthor{blanton05}) and $\rf = R-0.10$ (for
  \citeauthor{lin96,geller97,brown01}) have been used in ($c$).
  In ($d$), $\bj = g+0.25$  has been used to transform the LFs
  of \citeauthor{blanton05} and \citeauthor{bell03}, while $\bj = B_{\rm MGC}-0.13$ 
  was used on the LF of \citeauthor{driver05}.  The
  approximate absolute magnitude coverage of each survey is indicated by
  the extent of its curve.}
\label{fig:lf}
\end{figure*}

Figure~\ref{fig:lf} shows how the 6dFGS LFs compare to those of other
recent surveys.  In the \kb-band, the 6dFGS LF agrees remarkably well
with the earlier surveys using 2MASS photometry
\citep{cole01,kochanek01,bell03,eke05}. 
\citeauthor{eke05} find a faint-end slope which is flatter than other
recent LFs, including ours. The original \citeauthor{kochanek01}
magnitudes have had 0.05 and 0.135\,mags subtracted to convert from
isophotal to Kron magnitudes, and then to our total magnitudes. The
\citeauthor{cole01} Kron magnitudes have also had 0.135\,mag subtracted
to convert to total magnitudes. The larger sample and sky area of the
6dFGS provide $\sim 1.5$ to 2\,mags additional coverage at both the
bright and faint ends of the LF. We find that all previous surveys have
tended to slightly underestimate the bright end, the likely consequence
of smaller survey areas. \citeauthor{kochanek01} find a steeper
faint-end which they attribute to higher completeness and their brighter
apparent magnitude limit reducing sources of systematic error. However,
none of the other \kb-band surveys, including the 6dFGS, are as steep.
\citeauthor{bell03} claim to see evidence of a faint-end upturn in their
\kb-band LF and use an ordinary power law to fit the faint end beyond
$\mk - 5 \log h = -21$. The faint end slope of our ordinary Schechter
fit is a reasonable match to theirs over this range, and supports their
estimated surface brightness incompleteness correction.

The LFs in Fig.~\ref{fig:lf}($b$) show how the \citet{cole01} and
\citet{eke05} LFs compare with 6dFGS in \jb. Again, the \citeauthor{cole01} 
Kron magnitudes have been transformed to total magnitudes by subtracting 
$0.135$\,mag. As with \kb, the \jb-band luminosity functions are in close
agreement over those luminosities in common, although the 6dFGS
yields slightly higher densities of high-luminosity sources. The \citeauthor{eke05}
LF has a flatter \al\ and slightly higher normalisation than 6dFGS.

Figure~\ref{fig:lf}($c$) compares the 6dFGS \rf-band LF with those from
the Las Campanas Redshift Survey \citep[LCRS;][]{lin96} and Century
Survey with photographic \citep{geller97} and CCD photometry
\citep{brown01}. The $R$-band magnitudes of these surveys have been
shifted to \rf\ by subtracting $0.10$\,mag, following \citet{fukugita95}
for local galaxies of intermediate type. We also show the deep LF of
\citet{blanton05} for a special low-luminosity subset of SDSS, which we
have transformed to \rf\ from SDSS $r$ by subtracting $0.34$\,mag.  Our
\rf-band LF most closely agrees with the LCRS and SDSS surveys, and we
suspect that cosmic variance is largely responsible for differences with
the much smaller Century Survey. \citeauthor{blanton05} see evidence for
an upturn in the faint end and adopt a double-Schechter function to
match it. Although not as pronounced as \citeauthor{blanton05}, the
6dFGS LF faint end shows marginal evidence of a steeper rise fainter than 
$M - 5 \log h \sim -19$, although this is by no means conclusive.
Typically, a rising faint end is associated with dense cluster
environments \citep[and references therein]{driver04}. The
\citeauthor{blanton05} sample was deliberately chosen to be low
luminosity, and therefore, of comparatively small volume. It is possible
that the upturn they observe is the result of an overdensity in a nearby
portion of their sample volume.

In Fig.~\ref{fig:lf}($d$) we show how the 6dFGS \bj-band LFs compare to
those from the 2dFGRS \citep{norberg02}, Stromlo-APM \citep{loveday92}
and ESO Slice surveys \citep[ESP;][]{zucca97}. We have also transformed
the SDSS $g$-band LFs of \citet{bell03} and \citet{blanton05} by adding
$0.25$\,mag to match \bj, following \citeauthor{blanton05}. 
We also show the LF from the Millennium Galaxy Catalogue 
\citep[MGC;][]{driver05} which has been transformed from their blue passband
to ours through  $\bj = B_{\rm MGC}-0.13$. The LF of \citeauthor{zucca97} 
has been modified to take into account the cosmological model we have adopted.

Figure~\ref{fig:lf} shows that the 6dFGS faint end slope ($\al = -1.21$) is closest to
that of 2dFGRS and ESP, although the 6dFGS normalisation is lower than 
the others, possibly due to our non-use of evolution corrections (which are most
significant in bluer passbands). These magnitude corrections (e.g.\ 
\citeauthor{norberg02,bell03}) tend to lower the bright end end relative
to the faint end, and \citeauthor{blanton05} has demonstrated $\sim 20$
to 30\,percent variations are possible depending on the strength of the
applied correction. We note that the 6dFGS normalisation ($\log \ps =
-1.983$) most closely matches that of the \citeauthor{blanton05} SDSS
sample ($\log \ps = -1.928$), which (like the 6dFGS) was not corrected
for evolution. We do not find evidence for the steep faint end upturn
found by \citeauthor{blanton05}, despite having sufficient luminosity
coverage to test for it. We suspect a local overdensity is the likely
culprit for this feature of their data. Ultimately, subtle differences
are inevitable because of the way deeper but narrower surveys (SDSS,
2dFGRS, ESP) sample the very nearest large-scale structures compared to
the shallower but wider surveys such as Stromlo-APM and the 6dFGS.


\section{Luminosity Density}
\label{sec:lumdens}

\begin{figure}
\plotone{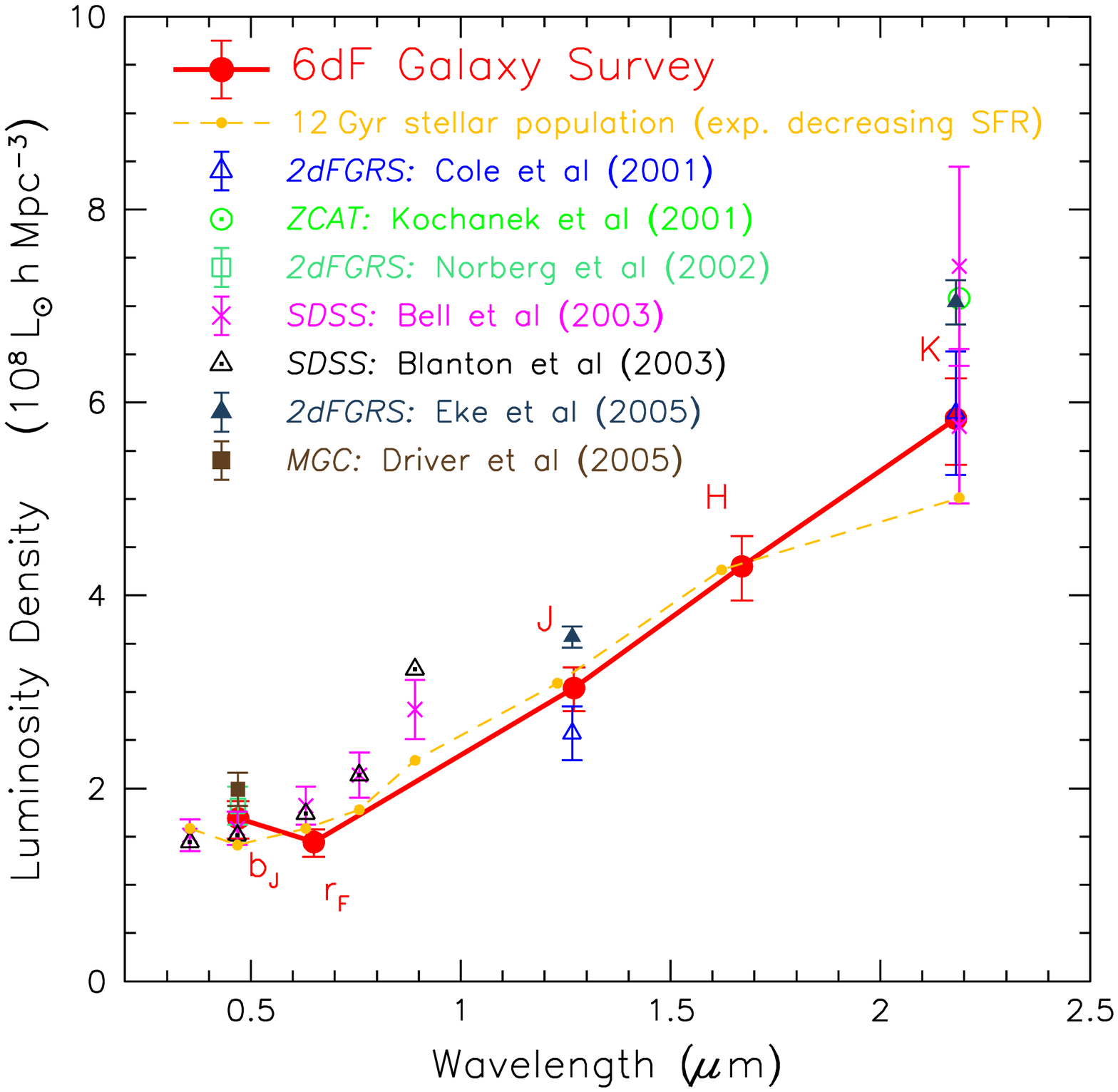}
\caption{
  Integrated luminosity density for 6dFGS across \brjhk\ in units of
  $10^8\,L_\odot\,h$\,Mpc$^{-3}$ (large red solid circles, solid red line).
  Comparitive values from other recent surveys are also shown.  Both
  \kb\ values of \citet{bell03} are included: one with and one without
  correction for low surface brightness selection effects. We have also
  reproduced the spectral energy distribution of \citet{bell03} for a
  12\,Gyr old stellar population with exponentially decreasing star
  formation rate, with $e$-folding time of 4\,Gyr (small solid circles,
  dashed line). }
\label{fig:lumdens}
\end{figure}

The total luminosity per unit volume, or luminosity density, $j$, is a
definitive observable of any galaxy population. It is the integral of
the luminosity function, weighted by luminosity, and is given in solar
luminosity units by
\begin{equation}
  j = \ps \cdot 10^{-0.4(M_*-M_\odot)} \cdot \Gamma(\al+2) .
\label{lumdens}
\end{equation}
Figure~\ref{fig:lumdens} shows the luminosity densities calculated from
Eqn.~(\ref{lumdens}) and the 6dFGS STY fit parameters in
Table~\ref{tab:schechtfits}.  We have used solar absolute magnitudes of
$M_\odot = (5.442, 4.447, 3.660, 3.319, 3.280)$ for \brjhk\ 
respectively, made available by
C.~Willmer\footnote{http://www.ucolick.org/$\sim$cnaw/sun.html}.  We
note that in the similar plot of \citet{bell03} (their Fig.~15) the \bj\ 
point for \cite{norberg02} is located $\sim 700$\,\AA\ blueward of its
true value.

Figure~\ref{fig:lumdens} also shows how the luminosity densities from
6dFGS compare to those of other recent surveys. Although much has been
made \citep[see e.g.][]{wright01,norberg02}, of $\sim 2 \times$
discrepancies between the early SDSS optical estimates \citep{blanton01}
and 2dFGRS in the NIR \citep{cole01}, the revised SDSS values of
\citet{blanton03}, corrected for cosmological evolution, have largely
ameliorated these differences. The 6dFGS values agree with most recent
estimates, finding a \kb-band value at the lower end of recent results
for this band. This is also the expectation of the spectral energy
distribution (SED) for a 12~Gyr old stellar population with a 4\,Gyr
$e$-folding exponentially decreasing star formation rate, \citep[dashed
line; after][]{bell03}. Models with more rapidly declining star
formation rates ($\sim 2$\,Gyr) can produce SEDs more luminous in \kb,
but at the expense of overestimating optical luminosities by factors of
$\sim 2$ or more.

While the luminosity density is a convenient quantity through which to
compare different surveys and infer a volume-averaged SED, it ultimately
relies on a fit to the LF and its extrapolation beyond accessible
luminosities. We find that the STY fits consistently overestimate the
integrated luminosity of the SWML method by $\sim 1$ to 3\,percent
across the observable range. The observed luminosity distribution
typically accounts for around 95\,percent of total luminosity, with the
shortfall due almost entirely to faint-end extrapolation. This is
comparable in size to the extra 2 to 5\,percent integrated luminosity
measured by the 6dFGS in reaching $\sim 1$ to 2\,mags fainter in the NIR
passbands.


\section{Summary}
\label{sec:summary}

We have presented luminosity functions in \khjrb\ from the 6dF Galaxy
Survey and derived luminosity densities from them. The 6dFGS target
samples are drawn from the 2MASS XSC and SuperCOSMOS catalogues and are
near-complete to limits of $(\kb,\hb,\jb,\rf,\bj) = (12.75, 13.00,
13.75, 15.60, 16.75)$. The samples used here represent around half the
final 6dFGS in terms of sky coverage and sample size. The 6dFGS is
already the largest NIR-selected redshift survey by more than an order
of magnitude in both coverage and sample size. Compared to the SDSS and
2dFGRS optical LFs it has around four times the sky coverage and half
the sample size.

The 6dFGS luminosity functions have been calculated using the \vmax,
STY and SWML estimators. The effects of magnitude and field-related
incompleteness have been characterised and taken into account.
Luminosity distances have been corrected for the effects of peculiar
velocity, which can amount to differences as large as $\sim \pm
0.5$\,mag for some supercluster members. The influence of photometric
errors on the shape of the LF has been taken into account for
\kb\hb\rf\bj, but was of little consequence for \jb.

These new 6dFGS LFs probe $\sim 1$ to 2\,mag fainter in absolute
magnitude than previous surveys in the NIR passbands and to comparable
limits in \rf\ and \bj. We obtain excellent agreement between the \vmax\ 
and SWML estimates of the LFs, but find that a Schechter function
(fitted by the optimal STY method) either under- or over-estimates these
values by as much as 15 to 40\,percent. While the formal uncertainties
on our STY fits are typically a few percent or less in $\log \ps$, \ms\ 
and \al, we show that, in comparison to the observed LF, a Schechter
function is unable to decline rapidly enough at the bright-end and
remain as flat at the faint end.

The 6dFGS results are generally in excellent agreement with other recent
LF determinations in the NIR. We do not find as steep a faint-end slope
as \citet{kochanek01}, and suspect this is due to their shallower depth
and subsequent brighter faint-end limit. The 6dFGS \rf-band LF most
closely matches those of the Las Campanas Redshift Survey \citep{lin96}
and SDSS \citep{blanton05}, although we find only marginal evidence for
the faint-end upturn claimed by \citeauthor{blanton05} In \bj, the 6dFGS
LF has a nearly identical faint end slope to those obtained by 2dFGRS
\citep{norberg02} and the ESO Slice Project \citep{zucca97}, although
the 6dFGS normalisation is closer to that found by
\citeauthor{blanton05}. Neither this survey nor 6dFGS used evolutionary
corrections. Furthermore, we see no evidence for a faint-end up-turn in
any of the 6dFGS LFs.

The luminosity densities derived in all five 6dFGS passbands concur with
most other recent measurements, and support a \kb-band value at the
lower range of recent values. This is consistent with the effective mean
galaxy spectral energy distribution being represented by an old stellar
population with moderately decreasing star formation rate.
 
\section*{Acknowledgements}

We would like to thank an anonymous referee for comments that 
improved the original manuscript in many ways. We are also grateful to
Val\'{e}rie de Lapparent for finding an error with Table 6 of the original
manuscript and drawing our attention to it.
We acknowledge the efforts of AAO staff at the UK Schmidt Telescope
whose ongoing dedication to 6dFGS has ensured its success. We are
similarly indebted to E.~Westra, M.~Williams, V.~Safouris, and S.~Prior
for their assistance to the project over time. We would like to thank
T.~Jarrett and M.~Read for their expert input on photometry issues
associated with the initial input catalogues. We acknowledge the
contributions of the 6dFGS Science Advisory Group: J.\ Huchra, T.\ 
Jarrett, O.\ Lahav, J.\ Lucey, G.~A.\ Mamon, Q.~A.\ Parker, D.\ Proust,
E.~M.\ Sadler, F.~G.\ Watson and K.\ Wakamatsu.

D.\ H.\ Jones is supported as a Research Associate by Australian
Research Council Discovery--Projects Grant (DP-0208876), administered by
the Australian National University.


\end{document}